\documentclass[aps,prb,twocolumn,showpacs,showkeys,superscriptaddress,groupedaddress]{revtex4}
\usepackage{inputenc}
\usepackage{natbib}
\usepackage{graphicx}  
\usepackage{epsfig}
\usepackage{float}
\usepackage{dcolumn}   
\usepackage{array}
\usepackage{bm}        
\usepackage{amssymb}   
\usepackage{amsmath}
\usepackage{amstext}
\usepackage{mathtools}
\usepackage [colorlinks=true,allcolors=blue]{hyperref}
\usepackage{footnote}
\usepackage{threeparttable}
\usepackage{siunitx} 
\begin{document}
\title{Strain induced Jahn-Teller distortions in BaFeO$_3$: A first-principles study}
\author{Imene Cherair}
\email [Corresponding author address:] {cherair.imene@gmail.com}
\affiliation{Theoretical Physics Laboratory, Faculty of Physics, USTHB,BP 32 El Alia, Bab Ezzouar, Algiers, Algeria.}
\author{Eric Bousquet}
\author{Michael Marcus Schmitt}
\affiliation{Theoretical Materials Physics, University of Li\`{e}ge(B5), B 4000 Li\`{e}ge, Belgium.}
\author{Nadia Iles}
\affiliation{Laboratory of Thin Films Physics and Materials for Electronics, Oran 1 University, Oran, Algeria}
\author{Abdelhafid Kellou}
\affiliation{Theoretical Physics Laboratory, Faculty of Physics, USTHB,BP 32 El Alia, Bab Ezzouar, Algiers, Algeria.}
\date{\today}
\begin{abstract}
The effect of epitaxial strain on structural, magnetic and electronic properties of BaFeO$_3$ perovskite oxide are investigated from first principles calculations, using the Density Functional Theory (DFT) plus the Hubbard approach (DFT+U) within the Generalized Gradient Approximation (GGA). Hybrid functional calculations, based on mixed exact Hartree-Fock (HF) and DFT exchange energy functionals, are also performed. For the ground state calculations,the DFT+U is found more suitable to describe the half metallic and ferromagnetic state of cubic BaFeO$_3$.
The possible occurence of oxygen octahedra rotations, Jahn-Teller distortions and charge orderings through biaxial strain are explored. 
The obtained results reveal that the Jahn-teller distortion is induced under tensile biaxial strain while the oxygen octahedra rotations and breathing are unusualy not observed. Then, the strained BaFeO$_3$ is considered as a particular Jahn-Teller distorted perovskite with exceptional properties when compared to CaFeO$_3$ and SrFeO$_3$. 
These findings lead to a strain engineering of the JT distortions in BaFeO$_3$, and thus for high fundamental and technological interests.

\end{abstract}
\maketitle

\section{introduction}
Iron perovskite oxides have raised a large interest in the last decades because of their particular electronic and magnetic properties.
A very appealing property of these compounds is to have a multiferroism behavior as often observed in BiFe$O_3$ and GaFe$O_3$\cite{Bahoosh2013}.
Unusual Fe$^{4+}$ (in high valence state) perovskites, such as cubic SrFeO$_3$ \cite{Takano1983,Okube2008-SrFeO3}, iron stabilize in the ${t_{2g}}^3{e_g}^1$ electronic configuration.
However, in CaFeO$_3$, a charge disproportionation is observed. This is provided by the 2Fe$^{4+}$ splitting into Fe$^{5+}$ and Fe$^{3+}$ oxidization states  \cite{Woodward2000-CaFeO3,Matsuno2002}.

BaFeO$_3$ belongs to these few Fe$^{4+}$ oxides where the tetravalent state of iron was confirmed very earlier\cite{Gallagher1965}. 
The physical properties of BaFeO$_3$ depend on oxygen content \cite{Hook1964, MacChesney1965,Grenier1989,Gonzalez1990}. Mori~\cite{Mori1966} demonstrated that BaFeO$_{3-y}$ forms in several phases: hexagonal 6H, triclinic, rhombohedral and tetragonal, depending on the oxygen off-stoichiometry. 
The 6H-hexagonal phase has attracted more attention because of interesting electronic and magnetic properties such as the charge disproportionation of Fe$^{+4}$~\cite{Morimoto2004,Mori2003}.

Stoichiometric BaFeO$_3$ was synthetized only recently by Hayashi \textit{et al}.~\cite{Hayashi2011} in 2011 where the cubic perovskite structure was found to be stable with a lattice constant $a=$3.971$ \si{\angstrom}$. 
The authors reported that a magnetic phase transition takes place at 111K from paramagnetic to A-type spin spiral order with a propagation vector along the [100] direction, the crystal remains metallic at all temperatures. 
By applying a small magnetic field of 0.3T, the spin sipral order is destroyed and a ferromagnetic (FM) phase is stabilized with a magnetic moment of 3.5 ${\mu}_B$ per Fe ion.
These features were examined theoretically by Li et al. \cite{Li22012}. They found that the A- and G-type helical orders in BaFeO$_3$ are almost degenerated to the ferromagnetic spin state with a small energy difference. 
A transition from helical order to ferromagnetic order is also found when the lattice constant is reduced \cite{Li2012}. 
Very recently, Rahman et al. \cite{Gul2016} demonstrated that the ground state of cubic BaFeO$_3$ is ferromagnetic. The transition from ferromagnetic to anti-ferromagnetic state is yielded only if displacements of Fe and O ions are considered. 
It was also shown from HAXPES and XAS measurements and from first-principle studies~\cite{Mizumaki2015} that the strong hybridization between Fe 3d and O 2p orbitals  are responsible for the negative charge transfer and metallicity.

Matsui et al. \cite{Matsui2002,Matsui2003} reported that BaFeO$_3$ films were found to be highly insulator with a weak ferromagnetism at T $>$ 300K. However, Callender et al. \cite{Callender2008} observed metallicity and strong ferromagnetism at $T_C$=235K that were attributed to the anionic deficiency. Both studies showed a pseudo cubic structures. In contrast, Taketani et al. \cite{Taketani2004} found the  BaFeO$_{y-3}$ films  have a tetragonal structure with an antiferromagnetic order. In addition, the decrease of oxygen vacancies increases the ferromagnetism in these films.
Chakraverty et al. \cite{Chakraverty2013}, synthesized fully oxidized  BaFeO$_3$ films by pulsed-laser deposition.They found that the films are ferromagnetic insulators with an optical gap of 1,8 eV. This result was  also confirmed by a X-ray study \cite{Tsuyama2015} which showed that the films have Fe$^{4+}$ valence state. 

Epitaxial strain in thin films technology is a well established technique to tune the perovskite properties and to induce new phases not observed in the unconstrained bulk crystals.\cite{Bousquet2008,Rabe2005,Bousquet2011,Lee2013}
Strain and interface effects can also be used together to observe new properties as in SrFe$O_3$/SrTiO$_3$ films \cite{Rondinelli2010}, in which oxygen octahedra rotations are induced in SrFeO$_3$ by SrTiO$_3$.
Motivated by these observations and the suggestion of Matsui \textit{et al.}~\cite{Matsui2002,Matsui2003} that BaFeO$_3$ may become a new magnetoelectric material, we propose in this work to study the effect of epitaxial strain on the properties of BaFeO$_3$. 
 Using first principles calculations, we investigate through biaxial strain the possible induced-occurence of a metal-insulator transition, magnetic phase transitions, oxygen octahedra rotations (OOR), Jahn-Teller (JT) distortions and charge orderings . 

\section{Computational details}
Our calculations have been performed using the Density Functional Theory, within the projected augmented wave (PAW) method \citealp{PAW2008}, as implemented in Abinit package \cite{Abinit2002,Abinit2005,Abinit2009,Abinit2016}.
The electronic exchange-correlation energy is described with the Generalized Gradient Approximation (GGA-PBE) \cite{PBE1996}. 
To improve the treatement of the localized Fe-3d electrons, the GGA plus Hubbard U (GGA+U) method has been employed \cite{DFT+U}, with the on-site coulomb effective $U$ ($U=$4eV).
 The $U$ values were tested from 0 to 8 eV. A good agreement of the bulk BaFeO$_3$ properties with the experimental data is found for U=4eV. This value was also used in previous theoretical works  \cite{Li2012,Ribeiro2013}.  
PAW atomic data files have been used from the JTH table (v0.1)~\cite{JTH} with the following valence states configurations: $(5s^25p^66s^2)$ for Ba, $(3s^23p^63d^74s^1)$ for Fe and $(2s^22p^6)$ for O.
The convergence on total energy of $10^{-4}Ha$ is reached for a plane wave energy cutoff of 20 Ha and PAW energy cutoff of 40 Ha. The Monkhorst-Pack scheme grid of $8\times8\times8$, $6\times6\times4$ and $4\times4\times4$ for the 5 atoms unit cells, 20 atoms supercell and 40 atoms supercell, respectively were sufficient to reach the above energy criterion.
Structural relaxations were carried out until the Helmann-Feynman forces become less than $10^{-6}$Ha/Bohr.
The effect of epitaxial stain was investigated through strained bulk calculations \cite{Dieguez2005} with a misfit strain defined as $\left( \frac{a-a_0}{a_0} \right) $, where $a_0$ is the calculated equilibrium lattice parameter of the five atoms unit cell, and a is the in-plane imposed cell parameter taken equal to  $\sqrt{2}a_0$. 
The space-group is determined with FINDSYM symmetry analysis software \cite{FinSym}. The Glazer notations \cite{Glazer1972} is used to describe the octahedral distortions.

 In order to check the accuracy of the pseudopotentials (PAW) calculations, additional calculations based on the Full Potential Linearized Augmented Plane Wave (FP-LAPW) method were performed, as implemented in Wien2k code \cite{Wien2k}.
The wave functions, electron charge densities, and potentials, inside the muffin-tin spheres are expanded in terms of the spherical harmonics up to angular momentum $l_{max}$ = 10, while the plane-waves are used in the interstitial region. The muffin-tin spheres radii $R_{MT}$ of 2.0, 1.9 and 1.2 a.u are used for  Ba, Fe and O respectively. Convergence is reached for 60 special k-points in the irreducible Brillouin zone and  $R_{MT}*K_{max}=8$, where $K_{max}$ is the plane wave cut-off.

To explore the effect of the Exchange and Correlation (XC) interaction on the electronic structure, several XC functionals are tested and compared to the GGA+U one. For this purpose, CRYSTAL14\cite{Dovesi2014} code based on  the linear combination of atomic orbitals method (LCAO) and atom centered Gaussian basis sets is also used. Hybrid functionals require  exact Hartree-Fock (HF) exchange calculations mixed with the exchange energy of some usual DFT exchange functional. 
In this work, three additionnal functionals are tested : the hybrid Becke-1 Wu-Cohen (B1WC)~\cite{Bilc2008}, Heyd-Scuseria-Ernzerhof HSE06~\cite{Heyd2003,Heyd2006} and HSEsol\cite{Schimka2011}.
 We tested the exact exchange mixing parameter of B1WC (16\%) for values going from 10\% to 25\%. 
The lattice parameters and atomic positions are fully optimized and relaxed taking into account carefully convergence tests. The atomic distortions are consistent with the space group $Pm\bar{3}m$ (5 atoms) and tetragonal $P4/mbm$ (10 atoms). 
To find the stable magnetic configurations, the 5 atoms unit cell of the cubic structure was extended to 20-atom supercell.
We used a Monkhorst-Pack scheme with $8\times8\times8$, $6\times6\times8$ and $6\times6\times4$ for the k-point sampling of the 5, 10 and 20 atoms unit cells respectively.
 The basis sets for Barium, Iron and Oxygen were taken from references \onlinecite{Mahmoud2014}, \onlinecite{Heifets2015} and \onlinecite{Erba2013} respectively.

\section{Results and discussions}
\subsection{Bulk properties}
The ferromagnetic (FM) state of the cubic BaFeO$_3$ (space group $Pm\bar{3}m$) is reviewed. The obtained lattice parameter, magnetic moment, bulk modulus and its pressure derivative are reported in the Table \ref{tab:opt-table}.  
Our results for full potential (Wien2k) and PAW (Abinit) methods are very close to each other as well as to the available experimental and theoretical ones. 
The best agreement with the experimental lattice parameter a is obtained for $U =$ 4 eV.
Our calculations show a magnetic moment of $\backsim$ 4 ${\mu}_B$ ($U=$ 4 eV). This result is consistent with the $t_{2g} 3 \uparrow e_g1 \uparrow$  electronic configuration of Fe$^{4+}$ and with recent works on BaFeO3 \cite{Mizumaki2015,CHERAIR2016}.

The LCAO calculations using CRYSTAL14 and hybrid functional HSE06 functional give the lattice constant of 3.95 $\si{\angstrom}$ as DFT+U (U = 0 \si{\electronvolt}) calculations.
The B1WC and HSEsol underestimate the lattice parameter a of about 0.04 $\si{\angstrom}$. The mixing  of the HF exchange  within the B1WC functional has not affected the lattice parameter. However, increasing this mixing parameter increases the magnetic moment of Fe from $\sim 3.6 \mu_B$ (10 \% mixing) to $\sim 3.9 \mu_B$  (25 \%  mixing). This means that the mixing parameter tends to localize the charge on  Fe-\textit{d} orbitals. 

 \begin{table}
 	\begin{threeparttable}
 		\caption{The optimized lattice parameter $a_0$, magnetic moment $\mu$ ,bulk modulus $B_0$ and its pressure derivative $B_p$ of cubic BaFeO$_3$.}
 		\label{tab:opt-table}
 		\centering
 		\begin{tabular*}{\linewidth}{@{\extracolsep{\fill}}llllll}
 			\toprule
 			& & $a_0(\si{\angstrom})$ & $B_0 (GPa)$ & $B_p$ & $\mu({\mu}_B)$ \\ \hline
 			Present & U=0eV & 3.950 & 132.6 & 4.77 & 3.60\\
 			work & U=4eV & 3.990 & 124.7 & 4.59 & 4.01 \\
 			\textit{\textbf{Abinit}}& U=8eV & 4.046 & 117.1 & 4.13 & 4.33 \\ \hline
 			Present & U=0eV & 3.955 & 136.1 & 4.92 & 3.66 \\
 			work & U=4eV & 3.995 & 132.3 & 4.50 & 4.02 \\
 			\textit{\textbf{Wien2k}}& U=8eV & 4.044 & 131.6 & 2.29 & 4.16 \\ \hline
 			& 10\% B1WC & 3.915 & \multicolumn{1}{c}{-}  & \multicolumn{1}{c}{-} & 3.62\\
 			Present & 16\% B1WC & 3.913 & \multicolumn{1}{c}{-}  & \multicolumn{1}{c}{-} & 3.76 \\
 			work& 25\% B1WC & 3.912 & \multicolumn{1}{c}{-} & \multicolumn{1}{c}{-} & 3.93 \\ 
 			\textit{\textbf{Crystal14}}  & HSE06 & 3.955 & \multicolumn{1}{c}{-} & \multicolumn{1}{c}{-} & 3.97 \\ 
 			& HSEsol & 3.916 & \multicolumn{1}{c}{-} & \multicolumn{1}{c}{-} & 3.91 \\ \hline            
 			Other  & Th.GGA & 3.956 \tnote{a} & 139.0 \tnote{a} & 4.30 \tnote{a} & 3.65 \tnote{b}  \\
 			works & Th.GGA+U & 4.026 \tnote{a} & 116.0 \tnote {a} & 4.50 \tnote{a} & 4.03 \tnote{b} \\
 			& Exp. & 3.971 \tnote{c} & 167.6 \tnote{d} & \multicolumn{1}{c}{-} & 3.50 \tnote{c} \\
 			\botrule
 		\end{tabular*}
 		\begin{tablenotes}
 			\footnotesize
 			\item[a] Ref.\bibpunct{[}{]}{;}{n}{}{} \cite{Hamdad2014}
 			\item[b] Ref. \cite{Mizumaki2015}
 			\item[c] Ref. \cite{Hayashi2011}
 			\item[d] Semiempirical value in Ref. \cite{Bulk-modulus} 
 		\end{tablenotes}
 	\end{threeparttable}   
 \end{table}

\begin{figure}
	\includegraphics[width=8cm,keepaspectratio=true]{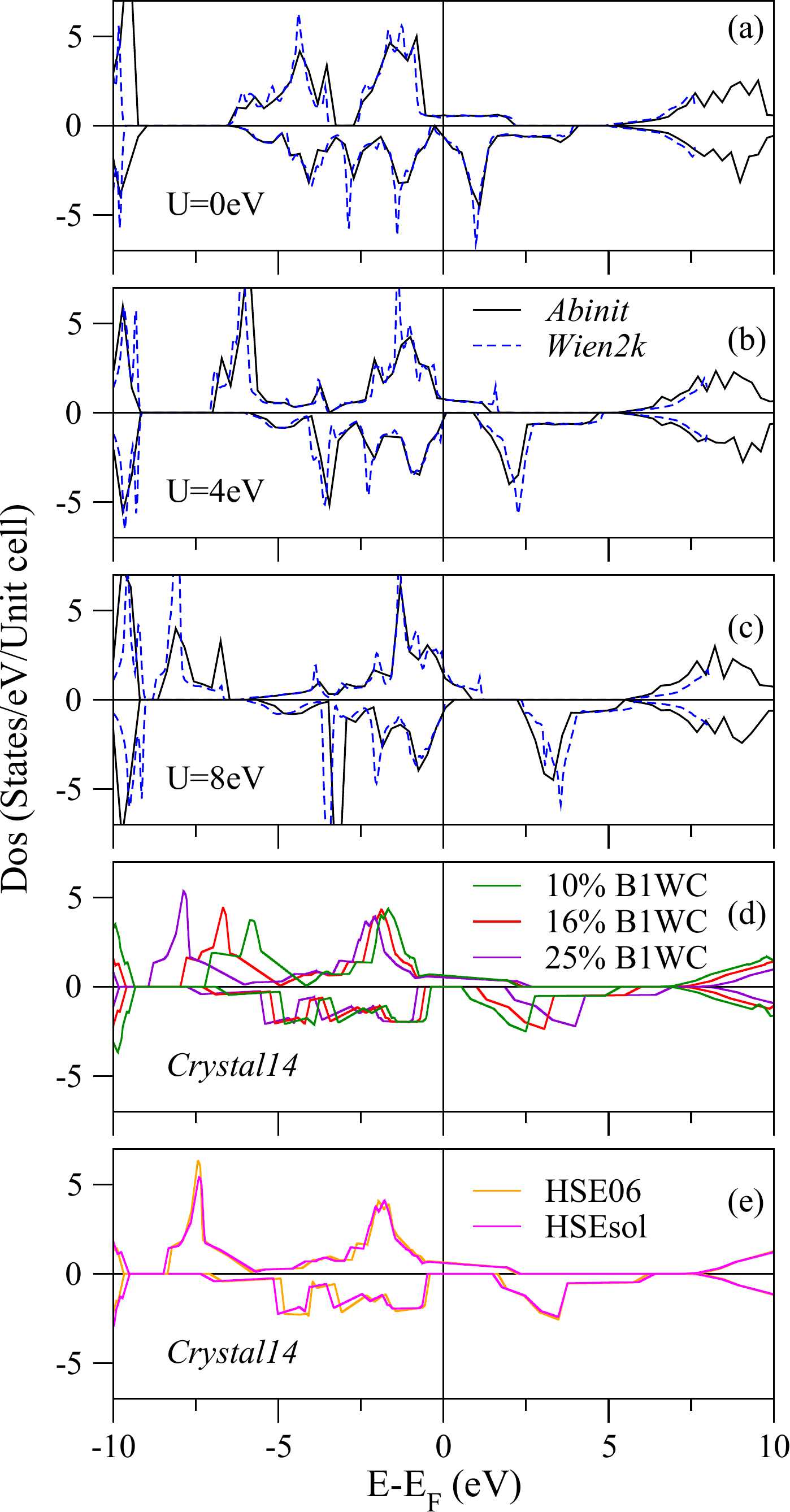}
	\caption{Total densities of states of the ferromagnetic cubic BaFeO$_3$ obtained by GGA and GGA+U using \textit{Abinit} and \textit{Wien2k} with (a) $U$=0eV, (b) $U$=4eV and (c) $U$=8eV, and \textit{Crystal14} (d), (e) with various Hybrid functionals .The Fermi level energy $E_F$ is located at 0 eV.} 
	\label{fig:fig1}
\end{figure}

The total density of state (DOS) obtained from the tested XC functionals are reported in the Fig \ref{fig:fig1}. The Fig \ref{fig:fig1} (a) - (c) show the obtained DFT$+U$ DOS using \textit{Wien2k} and \textit{Abinit} codes. 
BaFeO$_3$ is metallic for both up and down spin channels without the Hubbard U correction (Fig \ref{fig:fig1} a).
For $U=4$ eV (Fig \ref{fig:fig1} b), a gap is opened in the minority spin channel resulting in an half metallic character. 
A similar result is obtained for $U=8$ eV (Fig \ref{fig:fig1} c) but with an increased gap in the minority spin channel and a down shift in energy of the DOS around the Fermi level for the majority spin channel. These results are in agreement with the recent works of Mizumaki et al. \cite{Mizumaki2015} and Rahman et al. \cite{Gul2016}. 
All the hybrid functional calculations (Fig \ref{fig:fig1} d and e) give a half metallic behavior for the cubic structure of BaFeO$_3$. 
At a qualitative level, the calculated total DOSs  with the hybrid functionals B1WC (16\%),  HSE06 and HSEsol show the same trends as GGA$+U$ for $U=4$ eV.
Similarly, the Hubbard U value and the exact exchange mixing parameter of B1WC alter the band gap width in the minority spin channel (Fig \ref{fig:fig1} c).
While the HSE06 and HSEsol functionals give significantly different structural parameters, there is no sizable difference between them in the DOS (Fig \ref{fig:fig1} d). 
In order to find the most stable magnetic state of cubic BaFeO$_3$, several magnetic configurations were considered. The total energies of four magnetic orders (FM, A-AFM, C-AFM and G-AFM) were calculated. The FM state energy is taken as a reference. The corresponding differences in total energies are sum up in Table~\ref{tab:magnetic-order}. 
For all the used approaches (DFT$+U$ and hybrid functionals), the FM order is always the lowest energy state (all the energies relative to the FM one are positive in Table~\ref{tab:magnetic-order}) and the next lowest energy phase is the A-AFM. 
Our findings are in agreement  with the previous theoretical works of Li \textit{et al.}~\cite{LiTongwei2013}, Ribeiro \textit{et al.}~\cite{Ribeiro2013} and Cherair \textit{et al.}~\cite{CHERAIR2016}.
It can be noticed that the energy differences between the magnetic orders obtained with DFT$+U$ are more pronounced that those obtained with hybrid functionals. 

 \begin{table}
 \caption{Total energies (meV) of different AFM orders relative to the FM phase taken as the zero energy reference.}
 \label{tab:magnetic-order}
 \begin{tabular*}{\linewidth}{@{\extracolsep{\fill}}lcccc}
 \toprule
   & A-AFM & C-AFM & G-AFM \\ \hline
   $U$=0 eV & 375 & 831 & 1491 \\
   $U$=4 eV & 72 & 574 & 1416 \\
   $U$=8 eV & 568 & 895 & 1206 \\
   16 \% B1WC & 107 & 220  & 427 \\ 
   HSE06  & 125 & 236 & 461 \\
   HSEsol & 126 & 245 & 474 \\
   \botrule
   \end{tabular*}
 \end{table}
 
As it can be shown from the obtained bulk properties, the DFT+$U$ calculations with the PAW method ($Abinit$ code) provide the best agreement with previous experimental and theoretical ones. Accordingly, the PAW method is used in the next section to investigate the effect of structural distortions on the energy lowering of the cubic phase.
\subsection{Lattice distortion}

\subsubsection{Jahn Teller distortions}

\begin{table}
	\caption{Hybrid Functional lattice and magnetic results for tetragonal ferromagnetic $BaFeO_3$ (\textit{P4mbm}) and energy difference $\Delta E$ in \si{\milli\electronvolt} to the cubic relaxed phase calculated with the same hybrid exchange correlation functional.}
	\label{tab:P4mbm_hybrid}
	\begin{tabular*}{\linewidth}{@{\extracolsep{\fill}} l c c c c c c }
		\toprule 
 Functional & $a~(\si{\angstrom})$  & $c~(\si{\angstrom})$ & $a_0~(\si{\angstrom})$ & $Q_{JT}~(\si{\angstrom})$ & $\mu({\mu}_B)$ & $\Delta E(\si{\milli\electronvolt})$\\ 
		\hline
		 10 \% \textbf{B1WC} & 5.581 & 3.887  & 3.92 & 0.178 & 3.53 & -21.83 \\
	 16 \% \textbf{B1WC} & 5.591 & 3.873 & 3.91 & 0.205 & 3.62 & -35.58 \\
	  25 \% \textbf{B1WC} & 5.605 & 3.873 & 3.85 & 0.233 & 3.73 & -59.36  \\
	 \multicolumn{1}{c}{\textbf{HSE06}} & 5.629   & 3.931 & 3.96 & 0.121 & 3.90 & -9.31\\
	  \multicolumn{1}{c}{\textbf{HSEsol}} & 5.563  & 3.896 & 3.91 & 0.135 & 3.86  & -4.80 \\
		\botrule 
	\end{tabular*}
\end{table}	

\begin{figure}
	\includegraphics[width=3.4in]{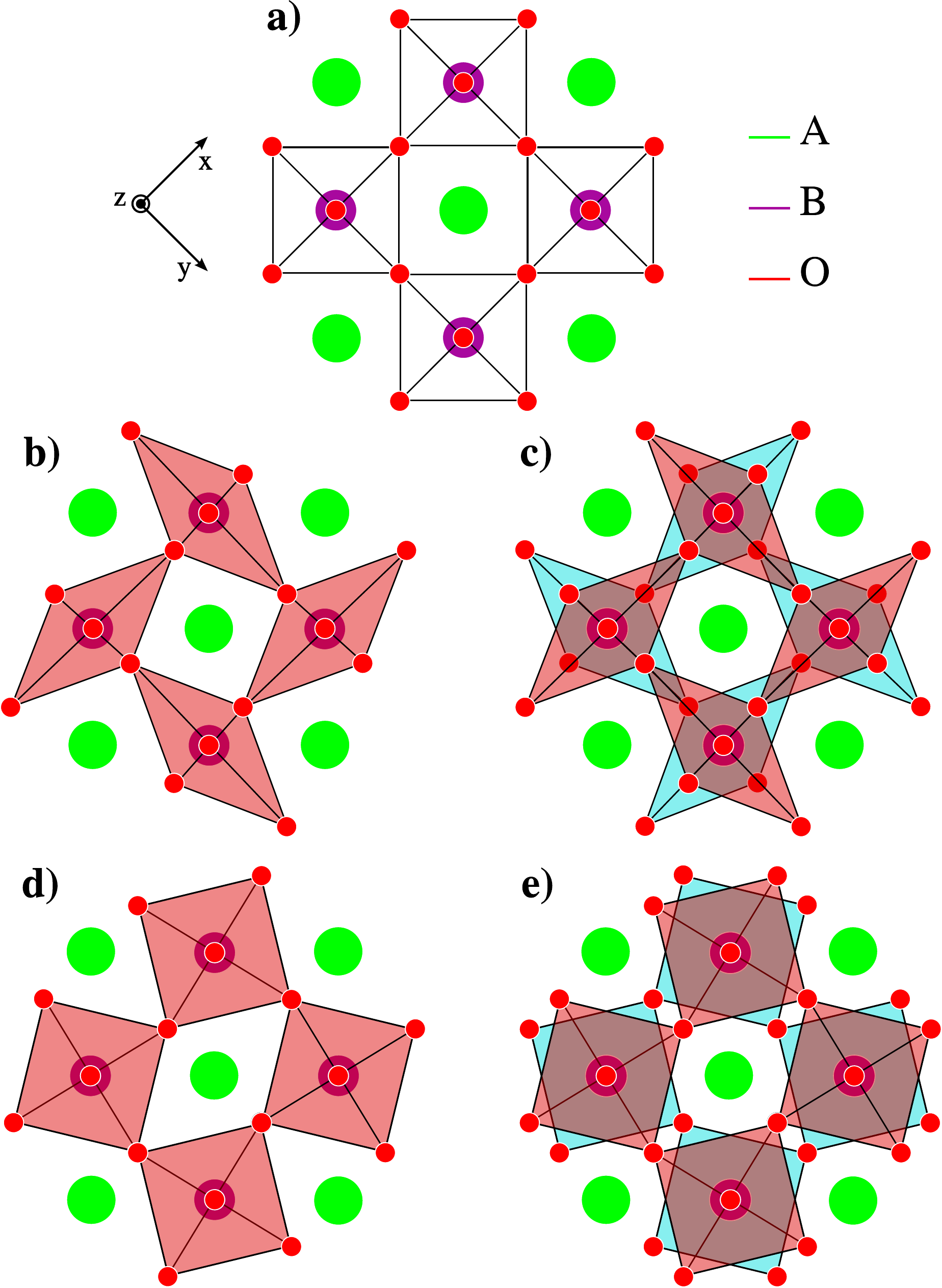}
	\caption{Schematic representation of Jahn-Teller and octahedral distortions within the perovskite ABO$_3$ structure. a) Undistorted ideal cubic perovskite structure. b) Jahn Teller at M point (JT-M). Octahedra in consecutive layers exhibit identical distortion. c) Jahn-Teller at R point (JT-R). Octahedra in consecutive layers exhibit opposite distortion. d) In phase rotation at M - point ($a^0a^0c^+$ in Glazer Notation). Consecutive Octahedra rotate about the same angle and direction. e) Out-of Phase rotation R point ($a^0a^0c^-$ in Glazer notation). Consecutive octahedra rotate about the same angle in opposite directions.}
	\label{fig:Rot_Distortions}
\end{figure}

In the following, the stability of the JT distortion with respect to the $U$ parameter within DFT$+U$ method is investigated. In a cubic octahedral environment and a high spin state,  the degeneracy of the $e_g$ orbitals of Fe$^{4+}$ is lifted by the occupancy of the fourth electron. This is accompanied by a change of symmetry from the cubic to the tetragonal one where the crystal distortion allows the energy lowering of the occupied $e_g$ orbital pointing toward the ligands. This fact induce an elongation or a contraction of one of the Fe-O bonds and the change of the two others through elasticity. This is the so-called Jahn Teller (JT) effect in crystals.
However, BaFeO$_3$ does not show any JT distortions while it has a high spin Fe$^{4+}$ configuration.
Two different JT distortions are defined: M-JT distortion from the M point (1/2,1/2,0.0) of the Brillouin zone, where each elongation/contraction of the Fe-O bonds are in phase along the $z$ direction (see Fig\ref{fig:Rot_Distortions}-b) and a R-JT distortion from the R point (1/2, 1/2, 1/2) of the Brillouin zone where each contraction and elongation of the Fe-O bonds are out-of-phase along the $z$ direction (see Fig \ref{fig:Rot_Distortions}-c).
When the M-JT distortion occcurs, it lowers the symmetry from cubic to tetragonal with the $P4/mbm$ space group (N\textsuperscript{o}127) and the R-JT to the $I4/mcm$ space group (N\textsuperscript{o}140).

\begin{figure*}
    \includegraphics[width=\textwidth,keepaspectratio=true]{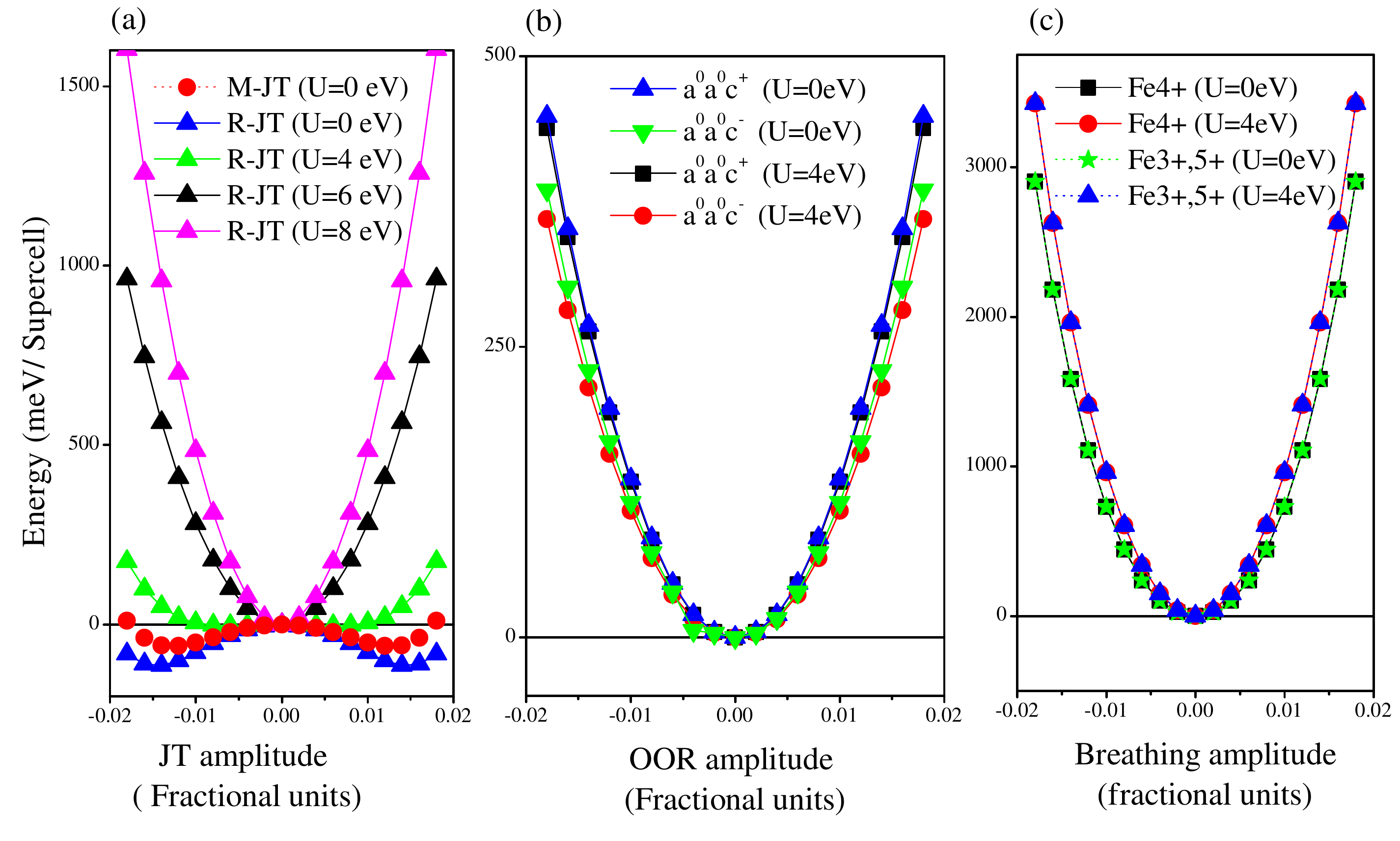}
    \caption{Energy wells with respect to the amplitude of (a) Jahn-Teller (JT)distortion (b) Oxygen octahedra rotations (OOR) and (c) Breathing-type distortion.The cubic energy is taken as zero} 
    \label{fig:distortion-curves}
\end{figure*}

The total energy variation with the amplitude of the JT distortions is reported in Fig (\ref{fig:distortion-curves}-a). For $U=0$ eV, the M- and R-JT patterns drive an energy double well and thus allow the energy lowering of the crystal with respect to the cubic reference (gain of energy of $\backsim$ 50 and 100 meV for the M- and R-JT, respectively).
Increasing the value of $U$ tends to reduce this gain of energy such as beyond $U=4$ eV the double wells are suppressed and are transformed into a single well for both M- and R-JT distortions. Consequently, it should be emphasized that  the $U$ parameter stabilizes the cubic symmetry against the JT distortions. To reproduce the cubic phase observed experimentally a $U$ parameter larger than 4 eV should be used for BaFeO$_3$.
This result goes well with Boukhvalov et al. findings \cite{Boukvalov2010}, who found that the JT distortion of BiMnO$_3$ is strongly dependent on the $U$ parameter; larger $U$ reduces the JT distortions. 
They attributed this effect to the increase of the Mn-O distances due to the $MnO_6$ octahedra expansion.

On Table \ref{tab:P4mbm_hybrid} are reported our results of the structural optimizations using CRYSTAL14 code with various hybrid exchange correlation functionals. We have used a supercell of 10 atoms restricted to the \textit{P4mbm} space group. 
According to our findings, for all hybrid functionals, the JT distortion lowers the energy with respect to the cubic reference phase. 
Hence, the hybrid functionals fail to predict the correct cubic ground-state of BaFeO$_3$. 
The B1WC functional gives the strongest energy gain (ten \si{\milli\electronvolt} depending on the HF/DFT exchange mixing). 
The amplitude of the JT distortion, the tetragonality and the magnetic moment on the Fe site are increased by increasing the HF exchange percentage. However, the HSE functionals gives smaller energy gain and the corresponding JT effect is energetically closer to the cubic phase ($\Delta E < \SI{10}{\milli\electronvolt}$).
Nonetheless, all the functionals fail to predict the undistorted cubic ground state even though HSE06 has been developed specifically for metallic systems \cite{Heyd2006}. 
Our findings are in agreement with the first-principles study using the HSE06 functional for the LaMO$_3$ 3 $d^0\rightarrow d^8$ (\textit{M = Sc - Cu}) compounds. In this study, HSE06 delivered accurate results except for the metallic paramagnets LaNiO$_3$ and LaCuO$_3$\cite{He2012}. 
The use of DFT$+U$ and a large enough $U$ is thus better adapted to the study of BaFeO$_3$ than the hybrid functionals.
We will thus focus our study with the DFT$+U$ approach in the rest of the paper.
It is intersting to note that in KCuF$_3$~\cite{DFT+U} the $U$ parameter is necessary to observe the JT distortion and in LaMnO$_3$~\cite{Sawada1997,mellan2015} the $U$ parameter has the tendency to enlarge the amplitude of the JT distortions. This is in opposition to what we observe in BaFeO$_3$.
In the case of BaFeO$_3$, the negative charge transfert is such that the increase of the $U$ parameter tends to increase the energetic interval between the Fe-d and O \textit{2p} orbitals. 
In other words, larger is the \textit{U} parameter less hybridized are Fe-\textit{d} and O-\textit{2p} orbitals and then less favorable is the JT instability which requires Fe-d and O-2p hybridization.
This can suggest that the covalency of the bonds in BaFeO$_3$ is low which is required to observe an undistorted cubic ground state.
  
\subsubsection{Oxygen octahedra rotations}
Oxygen octahedra rotations (OOR) are the most common distortions observed in perovskite crystals.~\cite{woodward1997,woodward1997b, Glazer1972}
We thus checked whether such octahedra rotations can be present in BaFeO$_3$. However, from the Goldshmidt factor \cite{goldschmidt1926} and from the experiment, it is unlikely that oxygen rotations appear. 
Accordingly, to confirm this fact, we studied the two most common OOR present in perovskites: The in-phase rotations of successive octahedra ($ a^0a^0c^+$ in the Galzer notation, see Fig \ref{fig:Rot_Distortions}-d for its representation), which appears at the M point of the Brillouin zone and the out-of-phase rotations ($a^0a^0c^-$), which comes from the R point (see Fig \ref{fig:Rot_Distortions}-e for its representation).
If present independently in one direction, these OOR lower the cubic symmetry to tetragonal one, with space groups $P4/mbm$ (127) and $I4/mcm$ (140) for the $ a^0a^0c^+$ and $ a^0a^0c^-$ cases, respectively. 

As for the JT distortion, we report in Fig (\ref{fig:distortion-curves}-b) the energy versus the $a^0a^0c^+$ and $ a^0a^0c^-$ distortions for $U=0$ and $U=4$ eV.
Contrary to the JT distortions, the OOR are always stable whatever the value of $U$ with steep single energy well.
As expected from the Goldshmit tolerance factor, we thus confirm that the OOR are not favorable in BaFeO$_3$.

\subsubsection{Breathing-type distortion}
\begin{figure}
    \centering
    \includegraphics[width=3.4in,keepaspectratio=true]{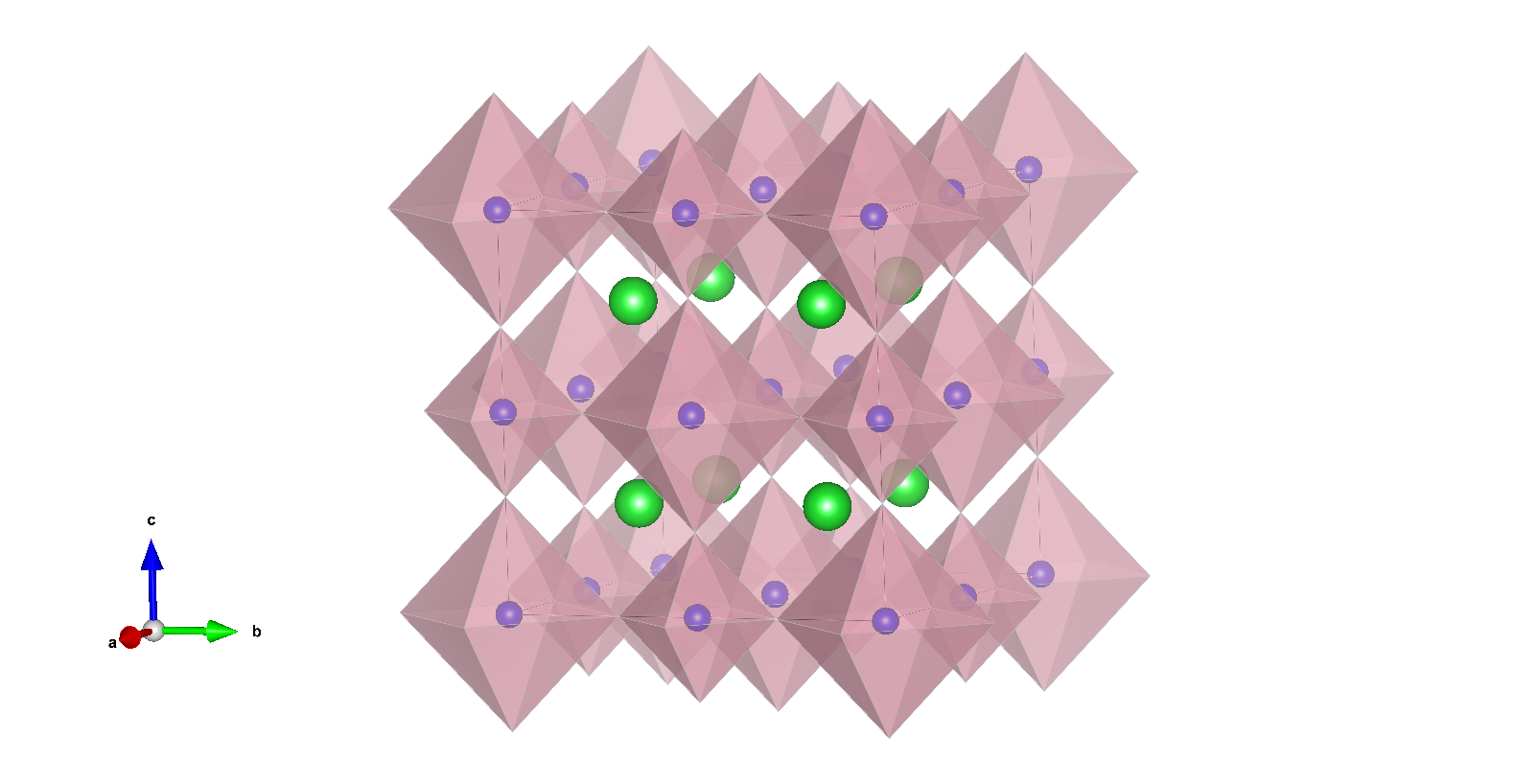}
    \caption{structural representation of breathing distortion.} 
    \label{fig:fig8}
    \end{figure}

The third type of perovskite distortions that we can analyze is the so called breathing of the oxygen octahedra.
This type of distortion is associated with a charge ordering of the transition metal cation where the differences of charge from site to site change the bond length with the surounding oxygens. 
This ends up with an alternating dilatation and contraction of the octahedra from site to site (See Fig \ref{fig:fig8} for a visualization of the breathing distortion).
The iron usually prefers to have an oxidation state of 2$+$ or 3$+$. 
In BaFeO$_3$ the Fe$^{4+}$ could have a tendency to split into the two oxidation states 3$+$ and 5$+$ (d$^5$ and d$^3$ orbital filling respectively) to lower the energy of the system and thus exhibiting a breathing distortion of the oxygen octahedra.
Such a distortion would lower the symmetry to the Fm$\bar{3}$m space group (number 225).\cite{Breathing}
This is not observed in experiments. To confirm this fact, we investigate the effect of the breathing distortion in BaFeO$_3$. We report the variation of the total energy versus several breathing distortions in Fig (\ref{fig:distortion-curves}-c) for $U=$0 and 4 eV.
We find that whatever the value of $U$ the breathing distortion forms a steep single well. Thus, we confirm that the breathing distortion is unlikely to be present in BaFeO$_3$. 
Our finding agrees with the recent work of Maznichenko \textit{et al.}\cite{Maznichenko2016}, who used DFT and XAS, XMCD experiments and they concluded that the oxidation of Fe in BaFeO$_3$ is $+$4 and it can change from +4 to +3 only if oxygen vacancies are present.

    \subsubsection{Polar distortions}
 BaFeO$_3$ is a metallic compound. Accordingly, we do not expect any polarization that can be present in the crystal. This is due to the full screening of the charges.
 Distortions that break the inversion center can be observed in metallic perovskite but it is related to steric or geometric effect,\cite{shi2014, garcia-castro2014}. However, these distortions are absent in BaFeO$_3$.
 As for the OOR and the breathing distortions, we do not find that polar-like distorions lower the energy of the cubic phase of BaFeO$_3$.
 The absence of all these distortions is consistent with the absence of phonon instabilities, already reported in the cubic phase of BaFeO$_3$.~\cite{CHERAIR2016}

 \subsection{Epitaxial strain effect}
 Epitaxial strain is nowadays a well known technique to tune the properties of perovskites. In some cases, this technique can also induce new phases absent in the unconstrained bulk crystal.
 We propose in the following to check whether epitaxial strain can induce the aforementioned structural instabilities discussed in the previous section.
 We thus analyze the stability of the JT, OOR and breathing distortions against the application of a biaxial strain.
 The biaxial strain can be simulated by fixing the two in-plane lattice parameters to simulate the epitaxial growth on a square lattice substrate where the out-of-plane lattice parameter is allowed to relax.\cite{Rabe2005} 
 We thus define the epitaxial strain as $\eta=\frac{a-a_0}{a_0}$ where $a_0$ is the fully relaxed unconstrained cubic lattice parameter and $a$ is the imposed cell parameter ($a=b$).
 We performed all of our calculation with strains ranging from $-4$\% to $+$4 \% and with GGA$+U=4$eV.

\begin{figure}
     \centering
     \includegraphics[width=3.4in,keepaspectratio=true]{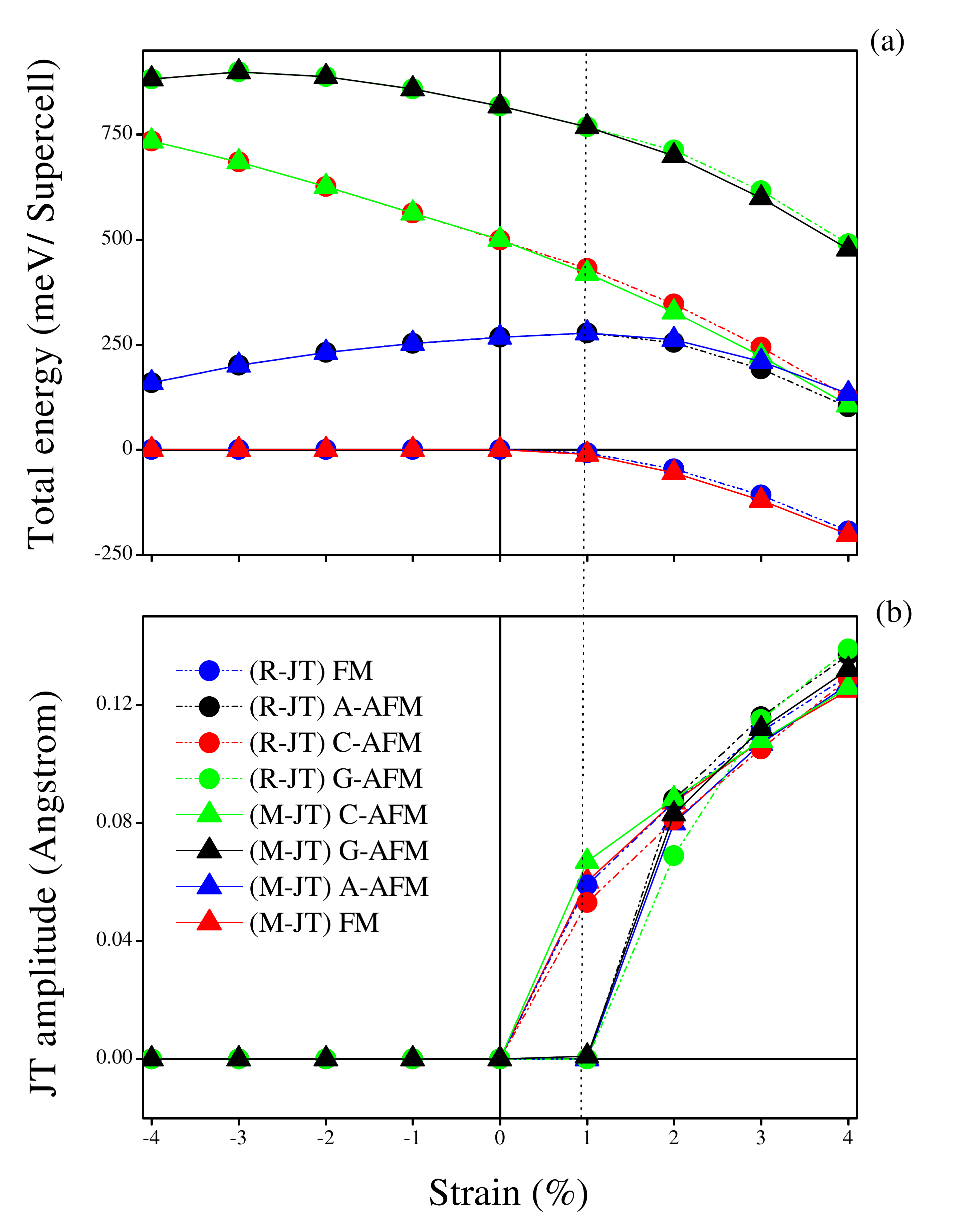}
     \caption{(a)Total energies in meV per (20-atoms supercell) of M-JT (triangles) and R-JT (circles) phases with respect to epitaxial strain for the FM, A-,C- and G-type magnetic orders. (b) The JT amplitude as function of epitaxial strain .} 
     \label{fig:strainUp}
     \end{figure}

\begin{figure}
 \centering
 \includegraphics[width=3.4in,keepaspectratio=true]{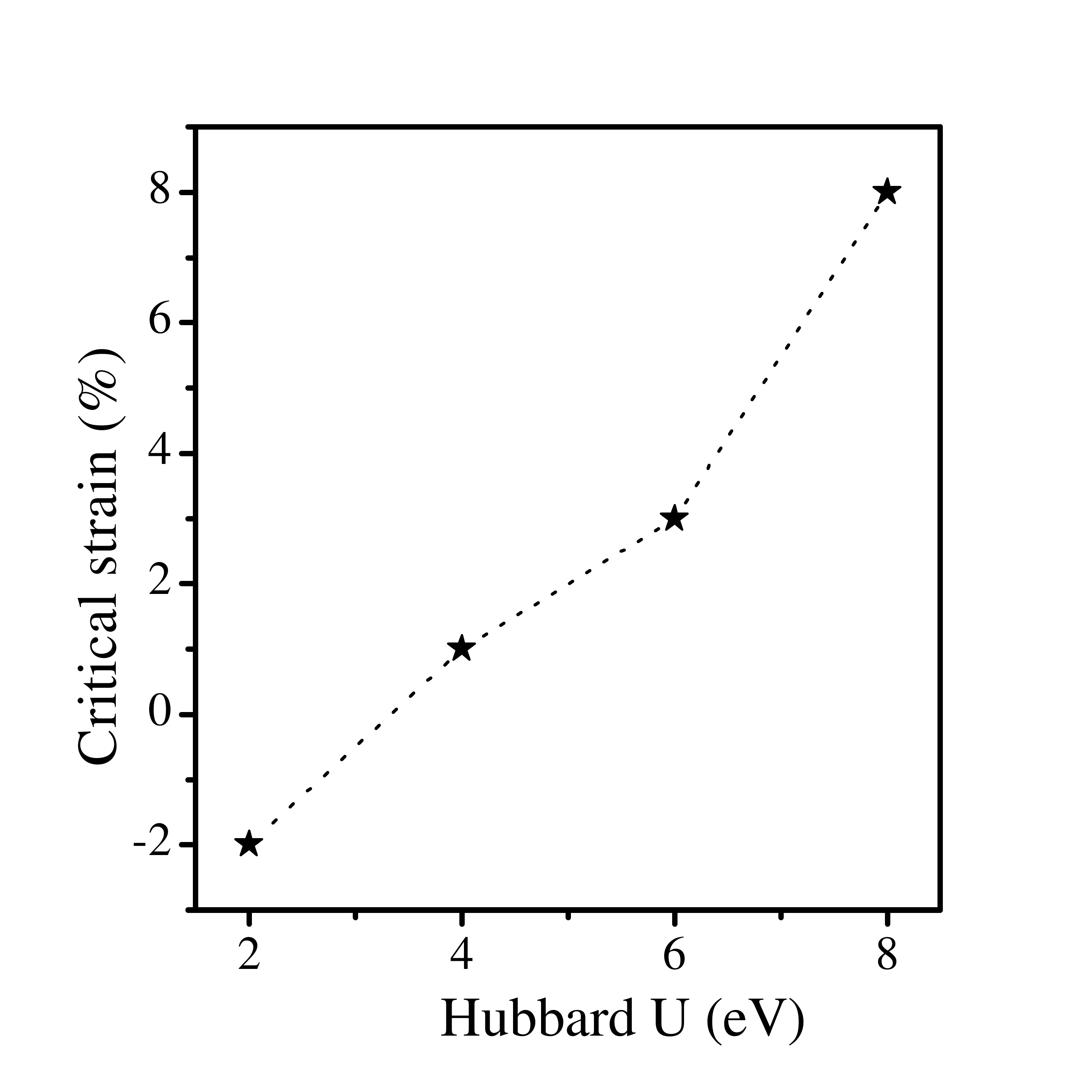}
 \caption{Variation of the critical strain to destabilise the JT distortions in BaFeO$_3$ as function of Hubbard U.} 
           \label{fig:critical-strain}
           \end{figure}

 In Fig (~\ref{fig:strainUp}-a), we report the total energy of both M-JT and R-JT phases versus epitaxial strain for the FM, G-, C-, and A-type magnetic orders (the undistorted FM phase. The atomic positions are considered in their high symmetry positions, but with the optimized $c$ cell parameter corresponding to a fixed $a=b$ in-plane cell parameters. 
 We can see from Fig (~\ref{fig:strainUp}-a) that the FM undistorted high symmetry reference is stable for compressive epitaxial strain going from $-$4\% to 0\%. This fact is also observed up to a tensile strain of $+$1\%. However, beyond $+$1\% of tensile strain, both FM R-JT and M-JT phases become lower in energy than the high symmetry reference structure. Thus, we can conclude that the tensile epitaxial strain stabilizes the JT distortions. When comparing the energy of the R-JT and M-JT phases under tensile strain larger than $+$1\%, we find that the M-JT has the lowest energy (the energy difference is 3, 8 ,12 and 9 meV/supercell at 1, 2, 3 and 4\% of tensile strain, respectively).
  Our present calculations show that beyond a critical tensile epitaxial strain BaFeO$_3$ exhibits a phase transition from the aristotype phase to an orbital ordering associated to the M-JT distortion.
  The critical value of $+$1\% (represented by a dashed vertical line in Fig (\ref{fig:strainUp}) is strongly dependent on the Hubbard value $U$. In Fig (\protect\ref{fig:critical-strain}), we plot the variation of the critical strain ($\eta_c$), necessary to observe the JT ordering, with the value of $U$. We can observe that $\eta_c$ increases with the value of $U$. We can also see that with $U$ = 2 eV, the critical strain is $-$2\%, which means that the JT appears even at 0\% strain and thus compressive epitaxial strain destroy the JT. Then, a value of $U\gtrsim 4$ eV is needed to describe the BaFe$O_3$ groundstate. In any case, qualitatively, we find that it exists a critical tensile epitaxial strain beyond which a JT phase transition is stabilized in BaFeO$_3$.\\
  For all the epitaxial strain values explored in our study we find that the FM phase is always the stable magnetic phase (in Fig (\ref{fig:strainUp}-a). The energies of the A-, C- and G-type phases are always higher in energy than the FM phase). We also notice that for compressive epitaxial strain, the energies of the A-type and the FM ordering are close. More interesting, we found that for stronger strain between -7\% and -8\%, the A-AFM state becomes lower in energy. 
  In Fig (\ref{fig:strainUp}-b), we report the amplitude of the JT-distortions for both R-JT and M-JT for the four magnetic orderings.
  Consistent with the Fig (\ref{fig:strainUp}-a), we find that the JT distortions become non zero at an epitaxial strain larger than $+$1\% for the FM phase. We also observe the same apparition of the JT distortions for A-,C- and G- type magnetic ordering at similar critical strain as for the FM phase.
  We also checked the possibility to induce OOR and breathing distortions at all values of strain but they never appear. This confirms that these distortions are not favorable under epitaxial strain in BaFeO$_3$.

\begin{figure}
\includegraphics[width=3.4in,keepaspectratio=true]{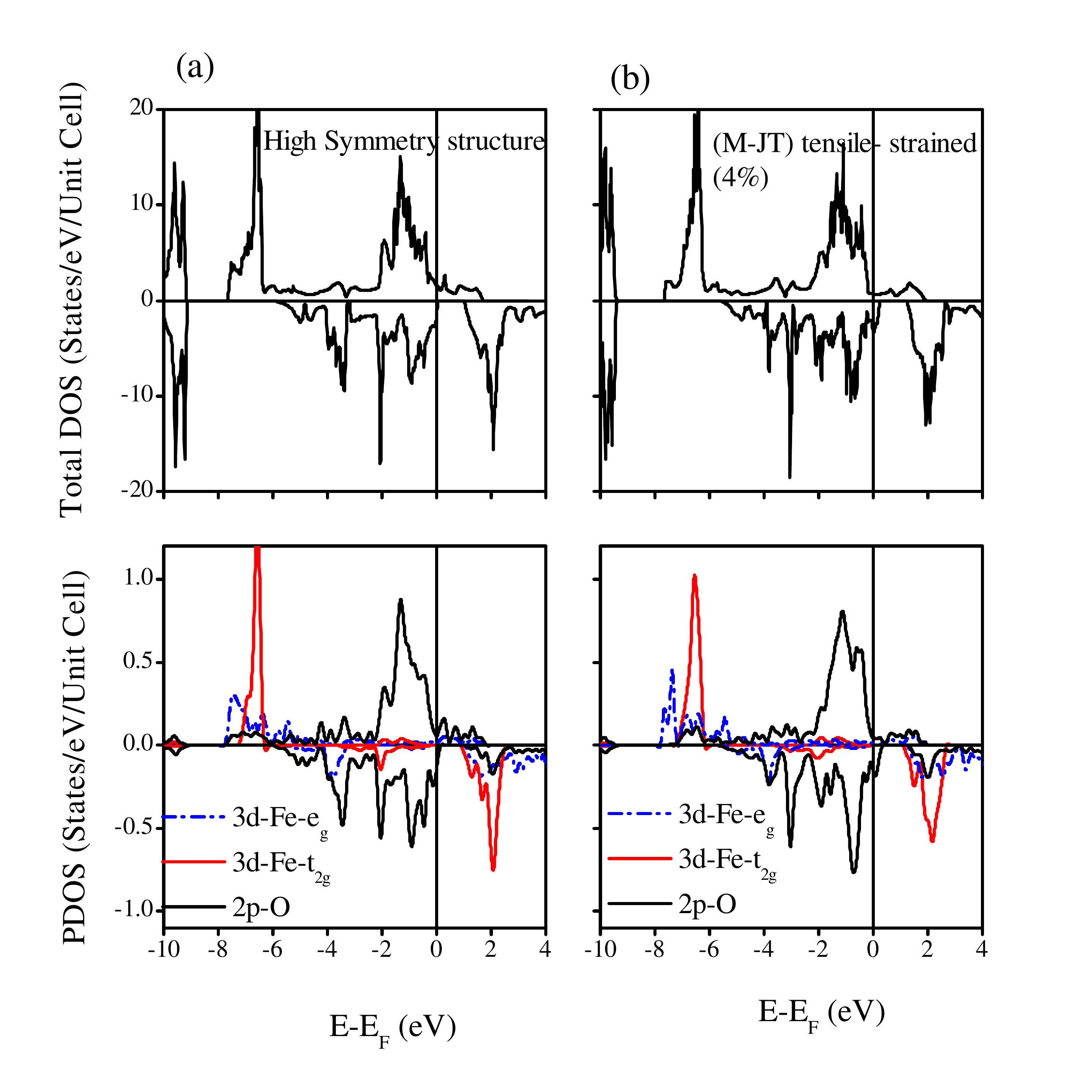}
\caption {Total (top graphs) and partial (down graphs) densities of state for (a) undistorted high-symmetry structure and (b) M-JT phase under $+4\%$ of strain. 
The Fermi level energy $E_F$ is placed at 0 eV.}
\label{fig:dos-strain}
\end{figure}
 
In Fig ~(\ref{fig:dos-strain}), we report the total and partial densities of states for the undistorted FM high-symmetry reference structure (column a in Figure~\ref{fig:dos-strain}) and the FM (M-JT) phase of BaFeO$_3$ under tensile strain of $+4\%$ value (column b in Figure~\ref{fig:dos-strain}) and thus within the GGA$+U$ ($U=$4eV) calculations.
We observe half metallicity in the case of the undistorted high symmetry structure (see Figure \ref{fig:dos-strain}-a) with a gap opened in the minority spin channel, similarly to the unstrained bulk.
In the presence of M-JT distortions (see Figure \ref{fig:dos-strain}-b), the the split of the Fe $3d$ states is broadened with respect to the undistorted case (Figure \ref{fig:dos-strain}-a) but no gap is opened.
The JT distortions narrow the $t_{2g}$ and $e_g$ bandwidth and shift the O 2p states above the Fermi level, resulting in a metallic behavior even for the minority spin.
The same observation in orthorhombic BaFeO$_3$ were reported by Rahman et al.\cite{Gul2016-strain}.
We note that the Fe 3d states are lower in energy than the O 2p states, which confirms a negative charge transfer in BaFeO$_3$ and might explain that the gap is not opened (the 2p orbitals are less localized than the 3d orbitals).
Recently, an metal-insulator transition in BaFe$O_3$ thin films is demonstrated experimentally by Tsuyama et al. \cite{Tsuyama2016}. According to our calculations, such a transition is not present in strained bulk crystal. The experimentally observed metal-insulator transition could originate from surface effects of the film or from oxygen vacancies, the later will directly affect the O 2p states present at the Fermi level and responsible for metallicity in BaFeO$_3$.
The possibility to control the electronic state modification offers a tremendous technological application, specially in the magnetic information manipulation and has been shown in strained magnetite \cite{Jeng2002,Friak2007} that exhibits a half-metal to metal transition. 
BaFeO$_3$ could be a new and interesting candidate for such applications due to its simple perovskite structure that can be combined easily in superlattices or grown as thin film.

\section{Conclusion}
In this paper, we have studied from DFT$+U$ and LCAO plus hybrid exchange correlation functional the properties of bulk and strained bulk BaFeO$_3$. 
The stability of BaFeO$_3$ against the most common structural distortions present in perovskites at the bulk level and under epitaxial strain have been investigated.
As for other metallic oxide systems, we found that BaFeO$_3$ is well described using the DFT$+U$ approach with $U\approx4\si{eV}$ rather than hybrid functionals. 
In agreement with experiments and previous reports, we find that the oxygen octahedra rotations and the breathing of the octahedra are strongly unfavorable in BaFeO$_3$. The high symmetry ferromagnetic half-metallic phase is more stable.
More interestingly, we found that a Jahn-Teller disortion can be induced in BaFeO$_3$ under tensile epitaxial strain.
We also observed that the Jahn-Teller distortions are strongly sensitive to the values of the $U$ parameter of the DFT$+U$ technique. This makes difficult to know the critical strain at which the Jahn-Teller distortions will appear. However, qualitatively, we can conclude that a Jahn-Teller distortion phase transition can be induced under tensile epitaxial strain.
Our study can motivate future experimental studies growing strained thin films of BaFeO$_3$ in order to induce a Jahn-Teller phase transition in the absence of other perovskite distortions.\cite{Mizokawa1999,Carpenter2009}
Consequently, the BaFeO$_3$ thin films can be used for magnetic manipulation \cite{Tsuyama2016} and in superlattices as a good candidate to provide robust multiferroic compounds \cite{Song2016}.
\bibliographystyle{apsrev4-1}

\begin{thebibliography}{72}%
\makeatletter
\providecommand \@ifxundefined [1]{%
 \@ifx{#1\undefined}
}%
\providecommand \@ifnum [1]{%
 \ifnum #1\expandafter \@firstoftwo
 \else \expandafter \@secondoftwo
 \fi
}%
\providecommand \@ifx [1]{%
 \ifx #1\expandafter \@firstoftwo
 \else \expandafter \@secondoftwo
 \fi
}%
\providecommand \natexlab [1]{#1}%
\providecommand \enquote  [1]{``#1''}%
\providecommand \bibnamefont  [1]{#1}%
\providecommand \bibfnamefont [1]{#1}%
\providecommand \citenamefont [1]{#1}%
\providecommand \href@noop [0]{\@secondoftwo}%
\providecommand \href [0]{\begingroup \@sanitize@url \@href}%
\providecommand \@href[1]{\@@startlink{#1}\@@href}%
\providecommand \@@href[1]{\endgroup#1\@@endlink}%
\providecommand \@sanitize@url [0]{\catcode `\\12\catcode `\$12\catcode
  `\&12\catcode `\#12\catcode `\^12\catcode `\_12\catcode `\%12\relax}%
\providecommand \@@startlink[1]{}%
\providecommand \@@endlink[0]{}%
\providecommand \url  [0]{\begingroup\@sanitize@url \@url }%
\providecommand \@url [1]{\endgroup\@href {#1}{\urlprefix }}%
\providecommand \urlprefix  [0]{URL }%
\providecommand \Eprint [0]{\href }%
\providecommand \doibase [0]{http://dx.doi.org/}%
\providecommand \selectlanguage [0]{\@gobble}%
\providecommand \bibinfo  [0]{\@secondoftwo}%
\providecommand \bibfield  [0]{\@secondoftwo}%
\providecommand \translation [1]{[#1]}%
\providecommand \BibitemOpen [0]{}%
\providecommand \bibitemStop [0]{}%
\providecommand \bibitemNoStop [0]{.\EOS\space}%
\providecommand \EOS [0]{\spacefactor3000\relax}%
\providecommand \BibitemShut  [1]{\csname bibitem#1\endcsname}%
\let\auto@bib@innerbib\@empty
\bibitem [{\citenamefont {Bahoosh}\ and\ \citenamefont
  {Wesselinowa}(2013)}]{Bahoosh2013}%
  \BibitemOpen
  \bibfield  {author} {\bibinfo {author} {\bibfnamefont {S.~G.}\ \bibnamefont
  {Bahoosh}}\ and\ \bibinfo {author} {\bibfnamefont {J.~M.}\ \bibnamefont
  {Wesselinowa}},\ }\href {\doibase 10.1063/1.4791586} {\bibfield  {journal}
  {\bibinfo  {journal} {Journal of Applied Physics}\ }\textbf {\bibinfo
  {volume} {113}},\ \bibinfo {eid} {063905} (\bibinfo {year} {2013}),\
  10.1063/1.4791586}\BibitemShut {NoStop}%
\bibitem [{\citenamefont {Takano}\ and\ \citenamefont
  {Takeda}(1983)}]{Takano1983}%
  \BibitemOpen
  \bibfield  {author} {\bibinfo {author} {\bibfnamefont {M.}~\bibnamefont
  {Takano}}\ and\ \bibinfo {author} {\bibfnamefont {Y.}~\bibnamefont
  {Takeda}},\ }\href {\doibase 2433/77050} {\bibfield  {journal} {\bibinfo
  {journal} {Bulletin of the Institute for Chemical Research, Kyoto
  University}\ }\textbf {\bibinfo {volume} {61}},\ \bibinfo {pages} {406}
  (\bibinfo {year} {1983})}\BibitemShut {NoStop}%
\bibitem [{\citenamefont {Okube}\ \emph {et~al.}(2008)\citenamefont {Okube},
  \citenamefont {Furukawa}, \citenamefont {Yoshiasa}, \citenamefont
  {Hashimoto}, \citenamefont {Sugahara},\ and\ \citenamefont
  {Nakatsuka}}]{Okube2008-SrFeO3}%
  \BibitemOpen
  \bibfield  {author} {\bibinfo {author} {\bibfnamefont {M.}~\bibnamefont
  {Okube}}, \bibinfo {author} {\bibfnamefont {Y.}~\bibnamefont {Furukawa}},
  \bibinfo {author} {\bibfnamefont {A.}~\bibnamefont {Yoshiasa}}, \bibinfo
  {author} {\bibfnamefont {T.}~\bibnamefont {Hashimoto}}, \bibinfo {author}
  {\bibfnamefont {M.}~\bibnamefont {Sugahara}}, \ and\ \bibinfo {author}
  {\bibfnamefont {A.}~\bibnamefont {Nakatsuka}},\ }\href
  {http://stacks.iop.org/1742-6596/121/i=9/a=092004} {\bibfield  {journal}
  {\bibinfo  {journal} {Journal of Physics: Conference Series}\ }\textbf
  {\bibinfo {volume} {121}},\ \bibinfo {pages} {092004} (\bibinfo {year}
  {2008})}\BibitemShut {NoStop}%
\bibitem [{\citenamefont {Woodward}\ \emph {et~al.}(2000)\citenamefont
  {Woodward}, \citenamefont {Cox}, \citenamefont {Moshopoulou}, \citenamefont
  {Sleight},\ and\ \citenamefont {Morimoto}}]{Woodward2000-CaFeO3}%
  \BibitemOpen
  \bibfield  {author} {\bibinfo {author} {\bibfnamefont {P.~M.}\ \bibnamefont
  {Woodward}}, \bibinfo {author} {\bibfnamefont {D.~E.}\ \bibnamefont {Cox}},
  \bibinfo {author} {\bibfnamefont {E.}~\bibnamefont {Moshopoulou}}, \bibinfo
  {author} {\bibfnamefont {A.~W.}\ \bibnamefont {Sleight}}, \ and\ \bibinfo
  {author} {\bibfnamefont {S.}~\bibnamefont {Morimoto}},\ }\href {\doibase
  10.1103/PhysRevB.62.844} {\bibfield  {journal} {\bibinfo  {journal} {Phys.
  Rev. B}\ }\textbf {\bibinfo {volume} {62}},\ \bibinfo {pages} {844} (\bibinfo
  {year} {2000})}\BibitemShut {NoStop}%
\bibitem [{\citenamefont {Matsuno}\ \emph {et~al.}(2002)\citenamefont
  {Matsuno}, \citenamefont {Mizokawa}, \citenamefont {Fujimori}, \citenamefont
  {Takeda}, \citenamefont {Kawasaki},\ and\ \citenamefont
  {Takano}}]{Matsuno2002}%
  \BibitemOpen
  \bibfield  {author} {\bibinfo {author} {\bibfnamefont {J.}~\bibnamefont
  {Matsuno}}, \bibinfo {author} {\bibfnamefont {T.}~\bibnamefont {Mizokawa}},
  \bibinfo {author} {\bibfnamefont {A.}~\bibnamefont {Fujimori}}, \bibinfo
  {author} {\bibfnamefont {Y.}~\bibnamefont {Takeda}}, \bibinfo {author}
  {\bibfnamefont {S.}~\bibnamefont {Kawasaki}}, \ and\ \bibinfo {author}
  {\bibfnamefont {M.}~\bibnamefont {Takano}},\ }\href {\doibase
  10.1103/PhysRevB.66.193103} {\bibfield  {journal} {\bibinfo  {journal} {Phys.
  Rev. B}\ }\textbf {\bibinfo {volume} {66}},\ \bibinfo {pages} {193103}
  (\bibinfo {year} {2002})}\BibitemShut {NoStop}%
\bibitem [{\citenamefont {Gallagher}\ \emph {et~al.}(1965)\citenamefont
  {Gallagher}, \citenamefont {MacChesney},\ and\ \citenamefont
  {Buchanan}}]{Gallagher1965}%
  \BibitemOpen
  \bibfield  {author} {\bibinfo {author} {\bibfnamefont {P.~K.}\ \bibnamefont
  {Gallagher}}, \bibinfo {author} {\bibfnamefont {J.~B.}\ \bibnamefont
  {MacChesney}}, \ and\ \bibinfo {author} {\bibfnamefont {D.~N.~E.}\
  \bibnamefont {Buchanan}},\ }\href {\doibase 10.1063/1.1696774} {\bibfield
  {journal} {\bibinfo  {journal} {The Journal of Chemical Physics}\ }\textbf
  {\bibinfo {volume} {43}},\ \bibinfo {pages} {516} (\bibinfo {year}
  {1965})}\BibitemShut {NoStop}%
\bibitem [{\citenamefont {Hook}(1964)}]{Hook1964}%
  \BibitemOpen
  \bibfield  {author} {\bibinfo {author} {\bibfnamefont {H.~J.~V.}\
  \bibnamefont {Hook}},\ }\href {\doibase 10.1021/j100794a041} {\bibfield
  {journal} {\bibinfo  {journal} {The Journal of Physical Chemistry}\ }\textbf
  {\bibinfo {volume} {68}},\ \bibinfo {pages} {3786} (\bibinfo {year}
  {1964})}\BibitemShut {NoStop}%
\bibitem [{\citenamefont {MacChesney}\ \emph {et~al.}(1965)\citenamefont
  {MacChesney}, \citenamefont {Potter}, \citenamefont {Sherwood},\ and\
  \citenamefont {Williams}}]{MacChesney1965}%
  \BibitemOpen
  \bibfield  {author} {\bibinfo {author} {\bibfnamefont {J.~B.}\ \bibnamefont
  {MacChesney}}, \bibinfo {author} {\bibfnamefont {J.~F.}\ \bibnamefont
  {Potter}}, \bibinfo {author} {\bibfnamefont {R.~C.}\ \bibnamefont
  {Sherwood}}, \ and\ \bibinfo {author} {\bibfnamefont {H.~J.}\ \bibnamefont
  {Williams}},\ }\href {\doibase 10.1063/1.1726393} {\bibfield  {journal}
  {\bibinfo  {journal} {The Journal of Chemical Physics}\ }\textbf {\bibinfo
  {volume} {43}},\ \bibinfo {pages} {3317} (\bibinfo {year}
  {1965})}\BibitemShut {NoStop}%
\bibitem [{\citenamefont {Grenier}\ \emph {et~al.}(1989)\citenamefont
  {Grenier}, \citenamefont {Wattiaux}, \citenamefont {Pouchard}, \citenamefont
  {Hagenmuller}, \citenamefont {Parras}, \citenamefont {Vallet}, \citenamefont
  {Calbet},\ and\ \citenamefont {Alario-Franco}}]{Grenier1989}%
  \BibitemOpen
  \bibfield  {author} {\bibinfo {author} {\bibfnamefont {J.-C.}\ \bibnamefont
  {Grenier}}, \bibinfo {author} {\bibfnamefont {A.}~\bibnamefont {Wattiaux}},
  \bibinfo {author} {\bibfnamefont {M.}~\bibnamefont {Pouchard}}, \bibinfo
  {author} {\bibfnamefont {P.}~\bibnamefont {Hagenmuller}}, \bibinfo {author}
  {\bibfnamefont {M.}~\bibnamefont {Parras}}, \bibinfo {author} {\bibfnamefont
  {M.}~\bibnamefont {Vallet}}, \bibinfo {author} {\bibfnamefont
  {J.}~\bibnamefont {Calbet}}, \ and\ \bibinfo {author} {\bibfnamefont
  {M.}~\bibnamefont {Alario-Franco}},\ }\href {\doibase
  10.1016/0022-4596(89)90025-X} {\bibfield  {journal} {\bibinfo  {journal}
  {Journal of Solid State Chemistry}\ }\textbf {\bibinfo {volume} {80}},\
  \bibinfo {pages} {6 } (\bibinfo {year} {1989})}\BibitemShut {NoStop}%
\bibitem [{\citenamefont {Gonzalez-Calbet}\ \emph {et~al.}(1990)\citenamefont
  {Gonzalez-Calbet}, \citenamefont {Parras}, \citenamefont {Vallet-Regi},\ and\
  \citenamefont {Grenier}}]{Gonzalez1990}%
  \BibitemOpen
  \bibfield  {author} {\bibinfo {author} {\bibfnamefont {J.}~\bibnamefont
  {Gonzalez-Calbet}}, \bibinfo {author} {\bibfnamefont {M.}~\bibnamefont
  {Parras}}, \bibinfo {author} {\bibfnamefont {M.}~\bibnamefont {Vallet-Regi}},
  \ and\ \bibinfo {author} {\bibfnamefont {J.}~\bibnamefont {Grenier}},\ }\href
  {\doibase 10.1016/0022-4596(90)90129-L} {\bibfield  {journal} {\bibinfo
  {journal} {Journal of Solid State Chemistry}\ }\textbf {\bibinfo {volume}
  {86}},\ \bibinfo {pages} {149 } (\bibinfo {year} {1990})}\BibitemShut
  {NoStop}%
\bibitem [{\citenamefont {MORI}(1966)}]{Mori1966}%
  \BibitemOpen
  \bibfield  {author} {\bibinfo {author} {\bibfnamefont {S.}~\bibnamefont
  {MORI}},\ }\href {\doibase 10.1111/j.1151-2916.1966.tb13176.x} {\bibfield
  {journal} {\bibinfo  {journal} {Journal of the American Ceramic Society}\
  }\textbf {\bibinfo {volume} {49}},\ \bibinfo {pages} {600} (\bibinfo {year}
  {1966})}\BibitemShut {NoStop}%
\bibitem [{\citenamefont {Morimoto}\ \emph {et~al.}(2004)\citenamefont
  {Morimoto}, \citenamefont {Kuzushita},\ and\ \citenamefont
  {Nasu}}]{Morimoto2004}%
  \BibitemOpen
  \bibfield  {author} {\bibinfo {author} {\bibfnamefont {S.}~\bibnamefont
  {Morimoto}}, \bibinfo {author} {\bibfnamefont {K.}~\bibnamefont {Kuzushita}},
  \ and\ \bibinfo {author} {\bibfnamefont {S.}~\bibnamefont {Nasu}},\ }\href
  {\doibase 10.1016/j.jmmm.2003.11.062} {\bibfield  {journal} {\bibinfo
  {journal} {Journal of Magnetism and Magnetic Materials}\ }\textbf {\bibinfo
  {volume} {272–276, Part 1}},\ \bibinfo {pages} {127 } (\bibinfo {year}
  {2004})},\ \bibinfo {note} {proceedings of the International Conference on
  Magnetism (ICM 2003)}\BibitemShut {NoStop}%
\bibitem [{\citenamefont {Mori}\ \emph {et~al.}(2003)\citenamefont {Mori},
  \citenamefont {Kamiyama}, \citenamefont {Kobayashi}, \citenamefont {Oikawa},
  \citenamefont {Otomo},\ and\ \citenamefont {Ikeda}}]{Mori2003}%
  \BibitemOpen
  \bibfield  {author} {\bibinfo {author} {\bibfnamefont {K.}~\bibnamefont
  {Mori}}, \bibinfo {author} {\bibfnamefont {T.}~\bibnamefont {Kamiyama}},
  \bibinfo {author} {\bibfnamefont {H.}~\bibnamefont {Kobayashi}}, \bibinfo
  {author} {\bibfnamefont {K.}~\bibnamefont {Oikawa}}, \bibinfo {author}
  {\bibfnamefont {T.}~\bibnamefont {Otomo}}, \ and\ \bibinfo {author}
  {\bibfnamefont {S.}~\bibnamefont {Ikeda}},\ }\href {\doibase
  10.1143/JPSJ.72.2024} {\bibfield  {journal} {\bibinfo  {journal} {Journal of
  the Physical Society of Japan}\ }\textbf {\bibinfo {volume} {72}},\ \bibinfo
  {pages} {2024} (\bibinfo {year} {2003})}\BibitemShut {NoStop}%
\bibitem [{\citenamefont {Hayashi}\ \emph {et~al.}(2011)\citenamefont
  {Hayashi}, \citenamefont {Yamamoto}, \citenamefont {Kageyama}, \citenamefont
  {Nishi}, \citenamefont {Watanabe}, \citenamefont {Kawakami}, \citenamefont
  {Matsushita}, \citenamefont {Fujimori},\ and\ \citenamefont
  {Takano}}]{Hayashi2011}%
  \BibitemOpen
  \bibfield  {author} {\bibinfo {author} {\bibfnamefont {N.}~\bibnamefont
  {Hayashi}}, \bibinfo {author} {\bibfnamefont {T.}~\bibnamefont {Yamamoto}},
  \bibinfo {author} {\bibfnamefont {H.}~\bibnamefont {Kageyama}}, \bibinfo
  {author} {\bibfnamefont {M.}~\bibnamefont {Nishi}}, \bibinfo {author}
  {\bibfnamefont {Y.}~\bibnamefont {Watanabe}}, \bibinfo {author}
  {\bibfnamefont {T.}~\bibnamefont {Kawakami}}, \bibinfo {author}
  {\bibfnamefont {Y.}~\bibnamefont {Matsushita}}, \bibinfo {author}
  {\bibfnamefont {A.}~\bibnamefont {Fujimori}}, \ and\ \bibinfo {author}
  {\bibfnamefont {M.}~\bibnamefont {Takano}},\ }\href {\doibase
  10.1002/anie.201105276} {\bibfield  {journal} {\bibinfo  {journal}
  {Angewandte Chemie International Edition}\ }\textbf {\bibinfo {volume}
  {50}},\ \bibinfo {pages} {12547} (\bibinfo {year} {2011})}\BibitemShut
  {NoStop}%
\bibitem [{\citenamefont {Li}\ \emph {et~al.}(2012{\natexlab{a}})\citenamefont
  {Li}, \citenamefont {Laskowski}, \citenamefont {Iitaka},\ and\ \citenamefont
  {Tohyama}}]{Li22012}%
  \BibitemOpen
  \bibfield  {author} {\bibinfo {author} {\bibfnamefont {Z.}~\bibnamefont
  {Li}}, \bibinfo {author} {\bibfnamefont {R.}~\bibnamefont {Laskowski}},
  \bibinfo {author} {\bibfnamefont {T.}~\bibnamefont {Iitaka}}, \ and\ \bibinfo
  {author} {\bibfnamefont {T.}~\bibnamefont {Tohyama}},\ }\href {\doibase
  10.1103/PhysRevB.85.134419} {\bibfield  {journal} {\bibinfo  {journal} {Phys.
  Rev. B}\ }\textbf {\bibinfo {volume} {85}},\ \bibinfo {pages} {134419}
  (\bibinfo {year} {2012}{\natexlab{a}})}\BibitemShut {NoStop}%
\bibitem [{\citenamefont {Li}\ \emph {et~al.}(2012{\natexlab{b}})\citenamefont
  {Li}, \citenamefont {Iitaka},\ and\ \citenamefont {Tohyama}}]{Li2012}%
  \BibitemOpen
  \bibfield  {author} {\bibinfo {author} {\bibfnamefont {Z.}~\bibnamefont
  {Li}}, \bibinfo {author} {\bibfnamefont {T.}~\bibnamefont {Iitaka}}, \ and\
  \bibinfo {author} {\bibfnamefont {T.}~\bibnamefont {Tohyama}},\ }\href
  {\doibase 10.1103/PhysRevB.86.094422} {\bibfield  {journal} {\bibinfo
  {journal} {Phys. Rev. B}\ }\textbf {\bibinfo {volume} {86}},\ \bibinfo
  {pages} {094422} (\bibinfo {year} {2012}{\natexlab{b}})}\BibitemShut
  {NoStop}%
\bibitem [{\citenamefont {Rahman}\ \emph {et~al.}(2016)\citenamefont {Rahman},
  \citenamefont {Morbec}, \citenamefont {Ferradás}, \citenamefont
  {García-Suárez},\ and\ \citenamefont {English}}]{Gul2016}%
  \BibitemOpen
  \bibfield  {author} {\bibinfo {author} {\bibfnamefont {G.}~\bibnamefont
  {Rahman}}, \bibinfo {author} {\bibfnamefont {J.~M.}\ \bibnamefont {Morbec}},
  \bibinfo {author} {\bibfnamefont {R.}~\bibnamefont {Ferradás}}, \bibinfo
  {author} {\bibfnamefont {V.~M.}\ \bibnamefont {García-Suárez}}, \ and\
  \bibinfo {author} {\bibfnamefont {N.~J.}\ \bibnamefont {English}},\ }\href
  {\doibase 10.1016/j.jmmm.2015.11.002} {\bibfield  {journal} {\bibinfo
  {journal} {Journal of Magnetism and Magnetic Materials}\ }\textbf {\bibinfo
  {volume} {401}},\ \bibinfo {pages} {1097 } (\bibinfo {year}
  {2016})}\BibitemShut {NoStop}%
\bibitem [{\citenamefont {Mizumaki}\ \emph {et~al.}(2015)\citenamefont
  {Mizumaki}, \citenamefont {Fujii}, \citenamefont {Yoshii}, \citenamefont
  {Hayashi}, \citenamefont {Saito}, \citenamefont {Shimakawa}, \citenamefont
  {Uozumi},\ and\ \citenamefont {Takano}}]{Mizumaki2015}%
  \BibitemOpen
  \bibfield  {author} {\bibinfo {author} {\bibfnamefont {M.}~\bibnamefont
  {Mizumaki}}, \bibinfo {author} {\bibfnamefont {H.}~\bibnamefont {Fujii}},
  \bibinfo {author} {\bibfnamefont {K.}~\bibnamefont {Yoshii}}, \bibinfo
  {author} {\bibfnamefont {N.}~\bibnamefont {Hayashi}}, \bibinfo {author}
  {\bibfnamefont {T.}~\bibnamefont {Saito}}, \bibinfo {author} {\bibfnamefont
  {Y.}~\bibnamefont {Shimakawa}}, \bibinfo {author} {\bibfnamefont
  {T.}~\bibnamefont {Uozumi}}, \ and\ \bibinfo {author} {\bibfnamefont
  {M.}~\bibnamefont {Takano}},\ }\href {\doibase 10.1002/pssc.201400252}
  {\bibfield  {journal} {\bibinfo  {journal} {physica status solidi (c)}\
  }\textbf {\bibinfo {volume} {12}},\ \bibinfo {pages} {818} (\bibinfo {year}
  {2015})}\BibitemShut {NoStop}%
\bibitem [{\citenamefont {Matsui}\ \emph {et~al.}(2002)\citenamefont {Matsui},
  \citenamefont {Tanaka}, \citenamefont {Fujimura}, \citenamefont {Ito},
  \citenamefont {Mabuchi},\ and\ \citenamefont {Morii}}]{Matsui2002}%
  \BibitemOpen
  \bibfield  {author} {\bibinfo {author} {\bibfnamefont {T.}~\bibnamefont
  {Matsui}}, \bibinfo {author} {\bibfnamefont {H.}~\bibnamefont {Tanaka}},
  \bibinfo {author} {\bibfnamefont {N.}~\bibnamefont {Fujimura}}, \bibinfo
  {author} {\bibfnamefont {T.}~\bibnamefont {Ito}}, \bibinfo {author}
  {\bibfnamefont {H.}~\bibnamefont {Mabuchi}}, \ and\ \bibinfo {author}
  {\bibfnamefont {K.}~\bibnamefont {Morii}},\ }\href {\doibase
  10.1063/1.1513213} {\bibfield  {journal} {\bibinfo  {journal} {Applied
  Physics Letters}\ }\textbf {\bibinfo {volume} {81}},\ \bibinfo {pages} {2764}
  (\bibinfo {year} {2002})}\BibitemShut {NoStop}%
\bibitem [{\citenamefont {{Matsui}}\ \emph {et~al.}(2003)\citenamefont
  {{Matsui}}, \citenamefont {{Taketani}}, \citenamefont {{Fujimura}},
  \citenamefont {{Ito}},\ and\ \citenamefont {{Morii}}}]{Matsui2003}%
  \BibitemOpen
  \bibfield  {author} {\bibinfo {author} {\bibfnamefont {T.}~\bibnamefont
  {{Matsui}}}, \bibinfo {author} {\bibfnamefont {E.}~\bibnamefont
  {{Taketani}}}, \bibinfo {author} {\bibfnamefont {N.}~\bibnamefont
  {{Fujimura}}}, \bibinfo {author} {\bibfnamefont {T.}~\bibnamefont {{Ito}}}, \
  and\ \bibinfo {author} {\bibfnamefont {K.}~\bibnamefont {{Morii}}},\ }\href
  {\doibase 10.1063/1.1556166} {\bibfield  {journal} {\bibinfo  {journal}
  {Journal of Applied Physics}\ }\textbf {\bibinfo {volume} {93}},\ \bibinfo
  {pages} {6993} (\bibinfo {year} {2003})}\BibitemShut {NoStop}%
\bibitem [{\citenamefont {Callender}\ \emph {et~al.}(2008)\citenamefont
  {Callender}, \citenamefont {Norton}, \citenamefont {Das}, \citenamefont
  {Hebard},\ and\ \citenamefont {Budai}}]{Callender2008}%
  \BibitemOpen
  \bibfield  {author} {\bibinfo {author} {\bibfnamefont {C.}~\bibnamefont
  {Callender}}, \bibinfo {author} {\bibfnamefont {D.~P.}\ \bibnamefont
  {Norton}}, \bibinfo {author} {\bibfnamefont {R.}~\bibnamefont {Das}},
  \bibinfo {author} {\bibfnamefont {A.~F.}\ \bibnamefont {Hebard}}, \ and\
  \bibinfo {author} {\bibfnamefont {J.~D.}\ \bibnamefont {Budai}},\ }\href
  {\doibase 10.1063/1.2832768} {\bibfield  {journal} {\bibinfo  {journal}
  {Applied Physics Letters}\ }\textbf {\bibinfo {volume} {92}},\ \bibinfo {eid}
  {012514} (\bibinfo {year} {2008}),\ 10.1063/1.2832768}\BibitemShut {NoStop}%
\bibitem [{\citenamefont {Taketani}\ \emph {et~al.}(2004)\citenamefont
  {Taketani}, \citenamefont {Matsui}, \citenamefont {Fujimura},\ and\
  \citenamefont {Morii}}]{Taketani2004}%
  \BibitemOpen
  \bibfield  {author} {\bibinfo {author} {\bibfnamefont {E.}~\bibnamefont
  {Taketani}}, \bibinfo {author} {\bibfnamefont {T.}~\bibnamefont {Matsui}},
  \bibinfo {author} {\bibfnamefont {N.}~\bibnamefont {Fujimura}}, \ and\
  \bibinfo {author} {\bibfnamefont {K.}~\bibnamefont {Morii}},\ }\href
  {\doibase 10.1109/TMAG.2004.830168} {\bibfield  {journal} {\bibinfo
  {journal} {IEEE transactions on magnetics}\ }\textbf {\bibinfo {volume}
  {40}},\ \bibinfo {pages} {2736} (\bibinfo {year} {2004})}\BibitemShut
  {NoStop}%
\bibitem [{\citenamefont {Chakraverty}\ \emph {et~al.}(2013)\citenamefont
  {Chakraverty}, \citenamefont {Matsuda}, \citenamefont {Ogawa}, \citenamefont
  {Wadati}, \citenamefont {Ikenaga}, \citenamefont {Kawasaki}, \citenamefont
  {Tokura},\ and\ \citenamefont {Hwang}}]{Chakraverty2013}%
  \BibitemOpen
  \bibfield  {author} {\bibinfo {author} {\bibfnamefont {S.}~\bibnamefont
  {Chakraverty}}, \bibinfo {author} {\bibfnamefont {T.}~\bibnamefont
  {Matsuda}}, \bibinfo {author} {\bibfnamefont {N.}~\bibnamefont {Ogawa}},
  \bibinfo {author} {\bibfnamefont {H.}~\bibnamefont {Wadati}}, \bibinfo
  {author} {\bibfnamefont {E.}~\bibnamefont {Ikenaga}}, \bibinfo {author}
  {\bibfnamefont {M.}~\bibnamefont {Kawasaki}}, \bibinfo {author}
  {\bibfnamefont {Y.}~\bibnamefont {Tokura}}, \ and\ \bibinfo {author}
  {\bibfnamefont {H.~Y.}\ \bibnamefont {Hwang}},\ }\href {\doibase
  10.1063/1.4824210} {\bibfield  {journal} {\bibinfo  {journal} {Applied
  Physics Letters}\ }\textbf {\bibinfo {volume} {103}},\ \bibinfo {eid}
  {142416} (\bibinfo {year} {2013}),\ 10.1063/1.4824210}\BibitemShut {NoStop}%
\bibitem [{\citenamefont {Tsuyama}\ \emph {et~al.}(2015)\citenamefont
  {Tsuyama}, \citenamefont {Matsuda}, \citenamefont {Chakraverty},
  \citenamefont {Okamoto}, \citenamefont {Ikenaga}, \citenamefont {Tanaka},
  \citenamefont {Mizokawa}, \citenamefont {Hwang}, \citenamefont {Tokura},\
  and\ \citenamefont {Wadati}}]{Tsuyama2015}%
  \BibitemOpen
  \bibfield  {author} {\bibinfo {author} {\bibfnamefont {T.}~\bibnamefont
  {Tsuyama}}, \bibinfo {author} {\bibfnamefont {T.}~\bibnamefont {Matsuda}},
  \bibinfo {author} {\bibfnamefont {S.}~\bibnamefont {Chakraverty}}, \bibinfo
  {author} {\bibfnamefont {J.}~\bibnamefont {Okamoto}}, \bibinfo {author}
  {\bibfnamefont {E.}~\bibnamefont {Ikenaga}}, \bibinfo {author} {\bibfnamefont
  {A.}~\bibnamefont {Tanaka}}, \bibinfo {author} {\bibfnamefont
  {T.}~\bibnamefont {Mizokawa}}, \bibinfo {author} {\bibfnamefont
  {H.}~\bibnamefont {Hwang}}, \bibinfo {author} {\bibfnamefont
  {Y.}~\bibnamefont {Tokura}}, \ and\ \bibinfo {author} {\bibfnamefont
  {H.}~\bibnamefont {Wadati}},\ }\href {\doibase 10.1103/PhysRevB.91.115101}
  {\bibfield  {journal} {\bibinfo  {journal} {Physical Review B}\ }\textbf
  {\bibinfo {volume} {91}},\ \bibinfo {pages} {115101} (\bibinfo {year}
  {2015})}\BibitemShut {NoStop}%
\bibitem [{\citenamefont {Bousquet}\ \emph {et~al.}(2008)\citenamefont
  {Bousquet}, \citenamefont {Dawber}, \citenamefont {Stucki}, \citenamefont
  {Lichtensteiger}, \citenamefont {Hermet}, \citenamefont {Gariglio},
  \citenamefont {Triscone},\ and\ \citenamefont {Ghosez}}]{Bousquet2008}%
  \BibitemOpen
  \bibfield  {author} {\bibinfo {author} {\bibfnamefont {E.}~\bibnamefont
  {Bousquet}}, \bibinfo {author} {\bibfnamefont {M.}~\bibnamefont {Dawber}},
  \bibinfo {author} {\bibfnamefont {N.}~\bibnamefont {Stucki}}, \bibinfo
  {author} {\bibfnamefont {C.}~\bibnamefont {Lichtensteiger}}, \bibinfo
  {author} {\bibfnamefont {P.}~\bibnamefont {Hermet}}, \bibinfo {author}
  {\bibfnamefont {S.}~\bibnamefont {Gariglio}}, \bibinfo {author}
  {\bibfnamefont {J.-M.}\ \bibnamefont {Triscone}}, \ and\ \bibinfo {author}
  {\bibfnamefont {P.}~\bibnamefont {Ghosez}},\ }\href {\doibase
  10.1038/nature06817} {\bibfield  {journal} {\bibinfo  {journal} {Nature}\
  }\textbf {\bibinfo {volume} {452}},\ \bibinfo {pages} {732} (\bibinfo {year}
  {2008})}\BibitemShut {NoStop}%
\bibitem [{\citenamefont {Rabe}(2005)}]{Rabe2005}%
  \BibitemOpen
  \bibfield  {author} {\bibinfo {author} {\bibfnamefont {K.~M.}\ \bibnamefont
  {Rabe}},\ }\href {\doibase 10.1016/j.cossms.2006.06.003} {\bibfield
  {journal} {\bibinfo  {journal} {Current Opinion in Solid State and Materials
  Science}\ }\textbf {\bibinfo {volume} {9}},\ \bibinfo {pages} {122} (\bibinfo
  {year} {2005})}\BibitemShut {NoStop}%
\bibitem [{\citenamefont {Bousquet}\ and\ \citenamefont
  {Spaldin}(2011)}]{Bousquet2011}%
  \BibitemOpen
  \bibfield  {author} {\bibinfo {author} {\bibfnamefont {E.}~\bibnamefont
  {Bousquet}}\ and\ \bibinfo {author} {\bibfnamefont {N.}~\bibnamefont
  {Spaldin}},\ }\href {\doibase 10.1103/PhysRevLett.107.197603} {\bibfield
  {journal} {\bibinfo  {journal} {Phys. Rev. Lett.}\ }\textbf {\bibinfo
  {volume} {107}},\ \bibinfo {pages} {197603} (\bibinfo {year}
  {2011})}\BibitemShut {NoStop}%
\bibitem [{\citenamefont {Lee}\ \emph {et~al.}(2013)\citenamefont {Lee},
  \citenamefont {Delaney}, \citenamefont {Bousquet}, \citenamefont {Spaldin},\
  and\ \citenamefont {Rabe}}]{Lee2013}%
  \BibitemOpen
  \bibfield  {author} {\bibinfo {author} {\bibfnamefont {J.~H.}\ \bibnamefont
  {Lee}}, \bibinfo {author} {\bibfnamefont {K.~T.}\ \bibnamefont {Delaney}},
  \bibinfo {author} {\bibfnamefont {E.}~\bibnamefont {Bousquet}}, \bibinfo
  {author} {\bibfnamefont {N.~A.}\ \bibnamefont {Spaldin}}, \ and\ \bibinfo
  {author} {\bibfnamefont {K.~M.}\ \bibnamefont {Rabe}},\ }\href {\doibase
  10.1103/PhysRevB.88.174426} {\bibfield  {journal} {\bibinfo  {journal} {Phys.
  Rev. B}\ }\textbf {\bibinfo {volume} {88}},\ \bibinfo {pages} {174426}
  (\bibinfo {year} {2013})}\BibitemShut {NoStop}%
\bibitem [{\citenamefont {Rondinelli}\ and\ \citenamefont
  {Spaldin}(2010)}]{Rondinelli2010}%
  \BibitemOpen
  \bibfield  {author} {\bibinfo {author} {\bibfnamefont {J.~M.}\ \bibnamefont
  {Rondinelli}}\ and\ \bibinfo {author} {\bibfnamefont {N.~A.}\ \bibnamefont
  {Spaldin}},\ }\href {\doibase 10.1103/PhysRevB.82.113402} {\bibfield
  {journal} {\bibinfo  {journal} {Phys. Rev. B}\ }\textbf {\bibinfo {volume}
  {82}},\ \bibinfo {pages} {113402} (\bibinfo {year} {2010})}\BibitemShut
  {NoStop}%
\bibitem [{\citenamefont {Torrent}\ \emph {et~al.}(2008)\citenamefont
  {Torrent}, \citenamefont {Jollet}, \citenamefont {Bottin}, \citenamefont
  {Z{\'e}rah},\ and\ \citenamefont {Gonze}}]{PAW2008}%
  \BibitemOpen
  \bibfield  {author} {\bibinfo {author} {\bibfnamefont {M.}~\bibnamefont
  {Torrent}}, \bibinfo {author} {\bibfnamefont {F.}~\bibnamefont {Jollet}},
  \bibinfo {author} {\bibfnamefont {F.}~\bibnamefont {Bottin}}, \bibinfo
  {author} {\bibfnamefont {G.}~\bibnamefont {Z{\'e}rah}}, \ and\ \bibinfo
  {author} {\bibfnamefont {X.}~\bibnamefont {Gonze}},\ }\href {\doibase
  10.1016/j.commatsci.2007.07.020} {\bibfield  {journal} {\bibinfo  {journal}
  {Computational Materials Science}\ }\textbf {\bibinfo {volume} {42}},\
  \bibinfo {pages} {337} (\bibinfo {year} {2008})}\BibitemShut {NoStop}%
\bibitem [{\citenamefont {Gonze}\ \emph {et~al.}(2002)\citenamefont {Gonze},
  \citenamefont {Beuken}, \citenamefont {Caracas}, \citenamefont {Detraux},
  \citenamefont {Fuchs}, \citenamefont {Rignanese}, \citenamefont {Sindic},
  \citenamefont {Verstraete}, \citenamefont {Zerah}, \citenamefont {Jollet}
  \emph {et~al.}}]{Abinit2002}%
  \BibitemOpen
  \bibfield  {author} {\bibinfo {author} {\bibfnamefont {X.}~\bibnamefont
  {Gonze}}, \bibinfo {author} {\bibfnamefont {J.-M.}\ \bibnamefont {Beuken}},
  \bibinfo {author} {\bibfnamefont {R.}~\bibnamefont {Caracas}}, \bibinfo
  {author} {\bibfnamefont {F.}~\bibnamefont {Detraux}}, \bibinfo {author}
  {\bibfnamefont {M.}~\bibnamefont {Fuchs}}, \bibinfo {author} {\bibfnamefont
  {G.-M.}\ \bibnamefont {Rignanese}}, \bibinfo {author} {\bibfnamefont
  {L.}~\bibnamefont {Sindic}}, \bibinfo {author} {\bibfnamefont
  {M.}~\bibnamefont {Verstraete}}, \bibinfo {author} {\bibfnamefont
  {G.}~\bibnamefont {Zerah}}, \bibinfo {author} {\bibfnamefont
  {F.}~\bibnamefont {Jollet}},  \emph {et~al.},\ }\href {\doibase
  10.1016/S0927-0256(02)00325-7} {\bibfield  {journal} {\bibinfo  {journal}
  {Computational Materials Science}\ }\textbf {\bibinfo {volume} {25}},\
  \bibinfo {pages} {478} (\bibinfo {year} {2002})}\BibitemShut {NoStop}%
\bibitem [{\citenamefont {Gonze}(2005)}]{Abinit2005}%
  \BibitemOpen
  \bibfield  {author} {\bibinfo {author} {\bibfnamefont {X.}~\bibnamefont
  {Gonze}},\ }\href {\doibase 10.1524/zkri.220.5.558.65066} {\bibfield
  {journal} {\bibinfo  {journal} {Zeitschrift f{\"u}r
  Kristallographie-Crystalline Materials}\ }\textbf {\bibinfo {volume} {220}},\
  \bibinfo {pages} {558} (\bibinfo {year} {2005})}\BibitemShut {NoStop}%
\bibitem [{\citenamefont {Gonze}\ \emph {et~al.}(2009)\citenamefont {Gonze},
  \citenamefont {Amadon}, \citenamefont {Anglade}, \citenamefont {Beuken},
  \citenamefont {Bottin}, \citenamefont {Boulanger}, \citenamefont {Bruneval},
  \citenamefont {Caliste}, \citenamefont {Caracas}, \citenamefont {Cote} \emph
  {et~al.}}]{Abinit2009}%
  \BibitemOpen
  \bibfield  {author} {\bibinfo {author} {\bibfnamefont {X.}~\bibnamefont
  {Gonze}}, \bibinfo {author} {\bibfnamefont {B.}~\bibnamefont {Amadon}},
  \bibinfo {author} {\bibfnamefont {P.-M.}\ \bibnamefont {Anglade}}, \bibinfo
  {author} {\bibfnamefont {J.-M.}\ \bibnamefont {Beuken}}, \bibinfo {author}
  {\bibfnamefont {F.}~\bibnamefont {Bottin}}, \bibinfo {author} {\bibfnamefont
  {P.}~\bibnamefont {Boulanger}}, \bibinfo {author} {\bibfnamefont
  {F.}~\bibnamefont {Bruneval}}, \bibinfo {author} {\bibfnamefont
  {D.}~\bibnamefont {Caliste}}, \bibinfo {author} {\bibfnamefont
  {R.}~\bibnamefont {Caracas}}, \bibinfo {author} {\bibfnamefont
  {M.}~\bibnamefont {Cote}},  \emph {et~al.},\ }\href {\doibase
  10.1016/j.cpc.2009.07.007} {\bibfield  {journal} {\bibinfo  {journal}
  {Computer Physics Communications}\ }\textbf {\bibinfo {volume} {180}},\
  \bibinfo {pages} {2582} (\bibinfo {year} {2009})}\BibitemShut {NoStop}%
\bibitem [{\citenamefont {Gonze}\ \emph {et~al.}(2016)\citenamefont {Gonze},
  \citenamefont {Jollet}, \citenamefont {Araujo}, \citenamefont {Adams},
  \citenamefont {Amadon}, \citenamefont {Applencourt}, \citenamefont {Audouze},
  \citenamefont {Beuken}, \citenamefont {Bieder}, \citenamefont {Bokhanchuk},
  \citenamefont {Bousquet}, \citenamefont {Bruneval}, \citenamefont {Caliste},
  \citenamefont {Côté}, \citenamefont {Dahm}, \citenamefont {Pieve},
  \citenamefont {Delaveau}, \citenamefont {Gennaro}, \citenamefont {Dorado},
  \citenamefont {Espejo}, \citenamefont {Geneste}, \citenamefont {Genovese},
  \citenamefont {Gerossier}, \citenamefont {Giantomassi}, \citenamefont
  {Gillet}, \citenamefont {Hamann}, \citenamefont {He}, \citenamefont {Jomard},
  \citenamefont {Janssen}, \citenamefont {Roux}, \citenamefont {Levitt},
  \citenamefont {Lherbier}, \citenamefont {Liu}, \citenamefont {Lukačević},
  \citenamefont {Martin}, \citenamefont {Martins}, \citenamefont {Oliveira},
  \citenamefont {Poncé}, \citenamefont {Pouillon}, \citenamefont {Rangel},
  \citenamefont {Rignanese}, \citenamefont {Romero}, \citenamefont {Rousseau},
  \citenamefont {Rubel}, \citenamefont {Shukri}, \citenamefont {Stankovski},
  \citenamefont {Torrent}, \citenamefont {Setten}, \citenamefont {Troeye},
  \citenamefont {Verstraete}, \citenamefont {Waroquiers}, \citenamefont
  {Wiktor}, \citenamefont {Xu}, \citenamefont {Zhou},\ and\ \citenamefont
  {Zwanziger}}]{Abinit2016}%
  \BibitemOpen
  \bibfield  {author} {\bibinfo {author} {\bibfnamefont {X.}~\bibnamefont
  {Gonze}}, \bibinfo {author} {\bibfnamefont {F.}~\bibnamefont {Jollet}},
  \bibinfo {author} {\bibfnamefont {F.~A.}\ \bibnamefont {Araujo}}, \bibinfo
  {author} {\bibfnamefont {D.}~\bibnamefont {Adams}}, \bibinfo {author}
  {\bibfnamefont {B.}~\bibnamefont {Amadon}}, \bibinfo {author} {\bibfnamefont
  {T.}~\bibnamefont {Applencourt}}, \bibinfo {author} {\bibfnamefont
  {C.}~\bibnamefont {Audouze}}, \bibinfo {author} {\bibfnamefont {J.-M.}\
  \bibnamefont {Beuken}}, \bibinfo {author} {\bibfnamefont {J.}~\bibnamefont
  {Bieder}}, \bibinfo {author} {\bibfnamefont {A.}~\bibnamefont {Bokhanchuk}},
  \bibinfo {author} {\bibfnamefont {E.}~\bibnamefont {Bousquet}}, \bibinfo
  {author} {\bibfnamefont {F.}~\bibnamefont {Bruneval}}, \bibinfo {author}
  {\bibfnamefont {D.}~\bibnamefont {Caliste}}, \bibinfo {author} {\bibfnamefont
  {M.}~\bibnamefont {Côté}}, \bibinfo {author} {\bibfnamefont
  {F.}~\bibnamefont {Dahm}}, \bibinfo {author} {\bibfnamefont {F.~D.}\
  \bibnamefont {Pieve}}, \bibinfo {author} {\bibfnamefont {M.}~\bibnamefont
  {Delaveau}}, \bibinfo {author} {\bibfnamefont {M.~D.}\ \bibnamefont
  {Gennaro}}, \bibinfo {author} {\bibfnamefont {B.}~\bibnamefont {Dorado}},
  \bibinfo {author} {\bibfnamefont {C.}~\bibnamefont {Espejo}}, \bibinfo
  {author} {\bibfnamefont {G.}~\bibnamefont {Geneste}}, \bibinfo {author}
  {\bibfnamefont {L.}~\bibnamefont {Genovese}}, \bibinfo {author}
  {\bibfnamefont {A.}~\bibnamefont {Gerossier}}, \bibinfo {author}
  {\bibfnamefont {M.}~\bibnamefont {Giantomassi}}, \bibinfo {author}
  {\bibfnamefont {Y.}~\bibnamefont {Gillet}}, \bibinfo {author} {\bibfnamefont
  {D.}~\bibnamefont {Hamann}}, \bibinfo {author} {\bibfnamefont
  {L.}~\bibnamefont {He}}, \bibinfo {author} {\bibfnamefont {G.}~\bibnamefont
  {Jomard}}, \bibinfo {author} {\bibfnamefont {J.~L.}\ \bibnamefont {Janssen}},
  \bibinfo {author} {\bibfnamefont {S.~L.}\ \bibnamefont {Roux}}, \bibinfo
  {author} {\bibfnamefont {A.}~\bibnamefont {Levitt}}, \bibinfo {author}
  {\bibfnamefont {A.}~\bibnamefont {Lherbier}}, \bibinfo {author}
  {\bibfnamefont {F.}~\bibnamefont {Liu}}, \bibinfo {author} {\bibfnamefont
  {I.}~\bibnamefont {Lukačević}}, \bibinfo {author} {\bibfnamefont
  {A.}~\bibnamefont {Martin}}, \bibinfo {author} {\bibfnamefont
  {C.}~\bibnamefont {Martins}}, \bibinfo {author} {\bibfnamefont
  {M.}~\bibnamefont {Oliveira}}, \bibinfo {author} {\bibfnamefont
  {S.}~\bibnamefont {Poncé}}, \bibinfo {author} {\bibfnamefont
  {Y.}~\bibnamefont {Pouillon}}, \bibinfo {author} {\bibfnamefont
  {T.}~\bibnamefont {Rangel}}, \bibinfo {author} {\bibfnamefont {G.-M.}\
  \bibnamefont {Rignanese}}, \bibinfo {author} {\bibfnamefont {A.}~\bibnamefont
  {Romero}}, \bibinfo {author} {\bibfnamefont {B.}~\bibnamefont {Rousseau}},
  \bibinfo {author} {\bibfnamefont {O.}~\bibnamefont {Rubel}}, \bibinfo
  {author} {\bibfnamefont {A.}~\bibnamefont {Shukri}}, \bibinfo {author}
  {\bibfnamefont {M.}~\bibnamefont {Stankovski}}, \bibinfo {author}
  {\bibfnamefont {M.}~\bibnamefont {Torrent}}, \bibinfo {author} {\bibfnamefont
  {M.~V.}\ \bibnamefont {Setten}}, \bibinfo {author} {\bibfnamefont {B.~V.}\
  \bibnamefont {Troeye}}, \bibinfo {author} {\bibfnamefont {M.}~\bibnamefont
  {Verstraete}}, \bibinfo {author} {\bibfnamefont {D.}~\bibnamefont
  {Waroquiers}}, \bibinfo {author} {\bibfnamefont {J.}~\bibnamefont {Wiktor}},
  \bibinfo {author} {\bibfnamefont {B.}~\bibnamefont {Xu}}, \bibinfo {author}
  {\bibfnamefont {A.}~\bibnamefont {Zhou}}, \ and\ \bibinfo {author}
  {\bibfnamefont {J.}~\bibnamefont {Zwanziger}},\ }\href {\doibase
  10.1016/j.cpc.2016.04.003} {\bibfield  {journal} {\bibinfo  {journal}
  {Computer Physics Communications}\ }\textbf {\bibinfo {volume} {205}},\
  \bibinfo {pages} {106 } (\bibinfo {year} {2016})}\BibitemShut {NoStop}%
\bibitem [{\citenamefont {Perdew}\ \emph {et~al.}(1996)\citenamefont {Perdew},
  \citenamefont {Burke},\ and\ \citenamefont {Ernzerhof}}]{PBE1996}%
  \BibitemOpen
  \bibfield  {author} {\bibinfo {author} {\bibfnamefont {J.~P.}\ \bibnamefont
  {Perdew}}, \bibinfo {author} {\bibfnamefont {K.}~\bibnamefont {Burke}}, \
  and\ \bibinfo {author} {\bibfnamefont {M.}~\bibnamefont {Ernzerhof}},\ }\href
  {\doibase 10.1103/PhysRevLett.77.3865} {\bibfield  {journal} {\bibinfo
  {journal} {Physical review letters}\ }\textbf {\bibinfo {volume} {77}},\
  \bibinfo {pages} {3865} (\bibinfo {year} {1996})}\BibitemShut {NoStop}%
\bibitem [{\citenamefont {Liechtenstein}\ \emph {et~al.}(1995)\citenamefont
  {Liechtenstein}, \citenamefont {Anisimov},\ and\ \citenamefont
  {Zaanen}}]{DFT+U}%
  \BibitemOpen
  \bibfield  {author} {\bibinfo {author} {\bibfnamefont {A.}~\bibnamefont
  {Liechtenstein}}, \bibinfo {author} {\bibfnamefont {V.}~\bibnamefont
  {Anisimov}}, \ and\ \bibinfo {author} {\bibfnamefont {J.}~\bibnamefont
  {Zaanen}},\ }\href {\doibase 10.1103/PhysRevB.52.R5467} {\bibfield  {journal}
  {\bibinfo  {journal} {Physical Review B}\ }\textbf {\bibinfo {volume} {52}},\
  \bibinfo {pages} {R5467} (\bibinfo {year} {1995})}\BibitemShut {NoStop}%
\bibitem [{\citenamefont {Ribeiro}\ \emph {et~al.}(2013)\citenamefont
  {Ribeiro}, \citenamefont {Godinho}, \citenamefont {Cardoso}, \citenamefont
  {Borges},\ and\ \citenamefont {Gasche}}]{Ribeiro2013}%
  \BibitemOpen
  \bibfield  {author} {\bibinfo {author} {\bibfnamefont {B.}~\bibnamefont
  {Ribeiro}}, \bibinfo {author} {\bibfnamefont {M.}~\bibnamefont {Godinho}},
  \bibinfo {author} {\bibfnamefont {C.}~\bibnamefont {Cardoso}}, \bibinfo
  {author} {\bibfnamefont {R.~P.}\ \bibnamefont {Borges}}, \ and\ \bibinfo
  {author} {\bibfnamefont {T.~P.}\ \bibnamefont {Gasche}},\ }\href {\doibase
  10.1063/1.4792664} {\bibfield  {journal} {\bibinfo  {journal} {Journal of
  Applied Physics}\ }\textbf {\bibinfo {volume} {113}},\ \bibinfo {eid}
  {083906} (\bibinfo {year} {2013}),\ 10.1063/1.4792664}\BibitemShut {NoStop}%
\bibitem [{\citenamefont {Jollet}\ \emph {et~al.}(2014)\citenamefont {Jollet},
  \citenamefont {Torrent},\ and\ \citenamefont {Holzwarth}}]{JTH}%
  \BibitemOpen
  \bibfield  {author} {\bibinfo {author} {\bibfnamefont {F.}~\bibnamefont
  {Jollet}}, \bibinfo {author} {\bibfnamefont {M.}~\bibnamefont {Torrent}}, \
  and\ \bibinfo {author} {\bibfnamefont {N.}~\bibnamefont {Holzwarth}},\ }\href
  {\doibase 10.1016/j.cpc.2013.12.023} {\bibfield  {journal} {\bibinfo
  {journal} {Computer Physics Communications}\ }\textbf {\bibinfo {volume}
  {185}},\ \bibinfo {pages} {1246 } (\bibinfo {year} {2014})}\BibitemShut
  {NoStop}%
\bibitem [{\citenamefont {Di\'eguez}\ \emph {et~al.}(2005)\citenamefont
  {Di\'eguez}, \citenamefont {Rabe},\ and\ \citenamefont
  {Vanderbilt}}]{Dieguez2005}%
  \BibitemOpen
  \bibfield  {author} {\bibinfo {author} {\bibfnamefont {O.}~\bibnamefont
  {Di\'eguez}}, \bibinfo {author} {\bibfnamefont {K.~M.}\ \bibnamefont {Rabe}},
  \ and\ \bibinfo {author} {\bibfnamefont {D.}~\bibnamefont {Vanderbilt}},\
  }\href {\doibase 10.1103/PhysRevB.72.144101} {\bibfield  {journal} {\bibinfo
  {journal} {Phys. Rev. B}\ }\textbf {\bibinfo {volume} {72}},\ \bibinfo
  {pages} {144101} (\bibinfo {year} {2005})}\BibitemShut {NoStop}%
\bibitem [{\citenamefont {Stokes}\ and\ \citenamefont {Hatch}(2005)}]{FinSym}%
  \BibitemOpen
  \bibfield  {author} {\bibinfo {author} {\bibfnamefont {H.~T.}\ \bibnamefont
  {Stokes}}\ and\ \bibinfo {author} {\bibfnamefont {D.~M.}\ \bibnamefont
  {Hatch}},\ }\href {\doibase 10.1063/1.4791586} {\bibfield  {journal}
  {\bibinfo  {journal} {Journal of Applied Crystallography}\ }\textbf {\bibinfo
  {volume} {38}},\ \bibinfo {pages} {237} (\bibinfo {year} {2005})}\BibitemShut
  {NoStop}%
\bibitem [{\citenamefont {Glazer}(1972)}]{Glazer1972}%
  \BibitemOpen
  \bibfield  {author} {\bibinfo {author} {\bibfnamefont {A.}~\bibnamefont
  {Glazer}},\ }\href {\doibase 10.1107/S0567740872007976} {\bibfield  {journal}
  {\bibinfo  {journal} {Acta Crystallographica Section B: Structural
  Crystallography and Crystal Chemistry}\ }\textbf {\bibinfo {volume} {28}},\
  \bibinfo {pages} {3384} (\bibinfo {year} {1972})}\BibitemShut {NoStop}%
\bibitem [{\citenamefont {Schwarz}\ \emph {et~al.}(2002)\citenamefont
  {Schwarz}, \citenamefont {Blaha},\ and\ \citenamefont {Madsen}}]{Wien2k}%
  \BibitemOpen
  \bibfield  {author} {\bibinfo {author} {\bibfnamefont {K.}~\bibnamefont
  {Schwarz}}, \bibinfo {author} {\bibfnamefont {P.}~\bibnamefont {Blaha}}, \
  and\ \bibinfo {author} {\bibfnamefont {G.}~\bibnamefont {Madsen}},\ }\href
  {\doibase 10.1016/S0010-4655(02)00206-0} {\bibfield  {journal} {\bibinfo
  {journal} {Computer Physics Communications}\ }\textbf {\bibinfo {volume}
  {147}},\ \bibinfo {pages} {71} (\bibinfo {year} {2002})}\BibitemShut
  {NoStop}%
\bibitem [{\citenamefont {Dovesi}\ \emph {et~al.}(2014)\citenamefont {Dovesi},
  \citenamefont {Orlando}, \citenamefont {Erba}, \citenamefont
  {Zicovich-Wilson}, \citenamefont {Civalleri}, \citenamefont {Casassa},
  \citenamefont {Maschio}, \citenamefont {Ferrabone}, \citenamefont
  {De~La~Pierre}, \citenamefont {D'Arco}, \citenamefont {Noël}, \citenamefont
  {Causà}, \citenamefont {Rérat},\ and\ \citenamefont
  {Kirtman}}]{Dovesi2014}%
  \BibitemOpen
  \bibfield  {author} {\bibinfo {author} {\bibfnamefont {R.}~\bibnamefont
  {Dovesi}}, \bibinfo {author} {\bibfnamefont {R.}~\bibnamefont {Orlando}},
  \bibinfo {author} {\bibfnamefont {A.}~\bibnamefont {Erba}}, \bibinfo {author}
  {\bibfnamefont {C.~M.}\ \bibnamefont {Zicovich-Wilson}}, \bibinfo {author}
  {\bibfnamefont {B.}~\bibnamefont {Civalleri}}, \bibinfo {author}
  {\bibfnamefont {S.}~\bibnamefont {Casassa}}, \bibinfo {author} {\bibfnamefont
  {L.}~\bibnamefont {Maschio}}, \bibinfo {author} {\bibfnamefont
  {M.}~\bibnamefont {Ferrabone}}, \bibinfo {author} {\bibfnamefont
  {M.}~\bibnamefont {De~La~Pierre}}, \bibinfo {author} {\bibfnamefont
  {P.}~\bibnamefont {D'Arco}}, \bibinfo {author} {\bibfnamefont
  {Y.}~\bibnamefont {Noël}}, \bibinfo {author} {\bibfnamefont
  {M.}~\bibnamefont {Causà}}, \bibinfo {author} {\bibfnamefont
  {M.}~\bibnamefont {Rérat}}, \ and\ \bibinfo {author} {\bibfnamefont
  {B.}~\bibnamefont {Kirtman}},\ }\href {\doibase 10.1002/qua.24658} {\bibfield
   {journal} {\bibinfo  {journal} {International Journal of Quantum Chemistry}\
  }\textbf {\bibinfo {volume} {114}},\ \bibinfo {pages} {1287} (\bibinfo {year}
  {2014})}\BibitemShut {NoStop}%
\bibitem [{\citenamefont {Bilc}\ \emph {et~al.}(2008)\citenamefont {Bilc},
  \citenamefont {Orlando}, \citenamefont {Shaltaf}, \citenamefont {Rignanese},
  \citenamefont {{\'I}{\~n}iguez},\ and\ \citenamefont {Ghosez}}]{Bilc2008}%
  \BibitemOpen
  \bibfield  {author} {\bibinfo {author} {\bibfnamefont {D.~I.}\ \bibnamefont
  {Bilc}}, \bibinfo {author} {\bibfnamefont {R.}~\bibnamefont {Orlando}},
  \bibinfo {author} {\bibfnamefont {R.}~\bibnamefont {Shaltaf}}, \bibinfo
  {author} {\bibfnamefont {G.-M.}\ \bibnamefont {Rignanese}}, \bibinfo {author}
  {\bibfnamefont {J.}~\bibnamefont {{\'I}{\~n}iguez}}, \ and\ \bibinfo {author}
  {\bibfnamefont {P.}~\bibnamefont {Ghosez}},\ }\href {\doibase
  10.1103/PhysRevB.77.165107} {\bibfield  {journal} {\bibinfo  {journal} {Phys.
  Rev. B}\ }\textbf {\bibinfo {volume} {77}},\ \bibinfo {pages} {165107}
  (\bibinfo {year} {2008})}\BibitemShut {NoStop}%
\bibitem [{\citenamefont {Heyd}\ \emph {et~al.}(2003)\citenamefont {Heyd},
  \citenamefont {Scuseria},\ and\ \citenamefont {Ernzerhof}}]{Heyd2003}%
  \BibitemOpen
  \bibfield  {author} {\bibinfo {author} {\bibfnamefont {J.}~\bibnamefont
  {Heyd}}, \bibinfo {author} {\bibfnamefont {G.~E.}\ \bibnamefont {Scuseria}},
  \ and\ \bibinfo {author} {\bibfnamefont {M.}~\bibnamefont {Ernzerhof}},\
  }\href {\doibase 10.1063/1.1564060} {\bibfield  {journal} {\bibinfo
  {journal} {The Journal of Chemical Physics}\ }\textbf {\bibinfo {volume}
  {118}},\ \bibinfo {pages} {8207} (\bibinfo {year} {2003})}\BibitemShut
  {NoStop}%
\bibitem [{\citenamefont {Heyd}\ \emph {et~al.}(2006)\citenamefont {Heyd},
  \citenamefont {Scuseria},\ and\ \citenamefont {Ernzerhof}}]{Heyd2006}%
  \BibitemOpen
  \bibfield  {author} {\bibinfo {author} {\bibfnamefont {J.}~\bibnamefont
  {Heyd}}, \bibinfo {author} {\bibfnamefont {G.~E.}\ \bibnamefont {Scuseria}},
  \ and\ \bibinfo {author} {\bibfnamefont {M.}~\bibnamefont {Ernzerhof}},\
  }\href {\doibase 10.1063} {\bibfield  {journal} {\bibinfo  {journal} {The
  Journal of Chemical Physics}\ }\textbf {\bibinfo {volume} {124}},\ \bibinfo
  {pages} {219906} (\bibinfo {year} {2006})}\BibitemShut {NoStop}%
\bibitem [{\citenamefont {Schimka}\ \emph {et~al.}(2011)\citenamefont
  {Schimka}, \citenamefont {Harl},\ and\ \citenamefont {Kresse}}]{Schimka2011}%
  \BibitemOpen
  \bibfield  {author} {\bibinfo {author} {\bibfnamefont {L.}~\bibnamefont
  {Schimka}}, \bibinfo {author} {\bibfnamefont {J.}~\bibnamefont {Harl}}, \
  and\ \bibinfo {author} {\bibfnamefont {G.}~\bibnamefont {Kresse}},\ }\href
  {\doibase 10.1063/1.3524336} {\bibfield  {journal} {\bibinfo  {journal} {The
  Journal of Chemical Physics}\ }\textbf {\bibinfo {volume} {134}},\ \bibinfo
  {pages} {024116} (\bibinfo {year} {2011})}\BibitemShut {NoStop}%
\bibitem [{\citenamefont {Mahmoud}\ \emph {et~al.}(2014)\citenamefont
  {Mahmoud}, \citenamefont {Erba}, \citenamefont {El-Kelany}, \citenamefont
  {R\'erat},\ and\ \citenamefont {Orlando}}]{Mahmoud2014}%
  \BibitemOpen
  \bibfield  {author} {\bibinfo {author} {\bibfnamefont {A.}~\bibnamefont
  {Mahmoud}}, \bibinfo {author} {\bibfnamefont {A.}~\bibnamefont {Erba}},
  \bibinfo {author} {\bibfnamefont {K.~E.}\ \bibnamefont {El-Kelany}}, \bibinfo
  {author} {\bibfnamefont {M.}~\bibnamefont {R\'erat}}, \ and\ \bibinfo
  {author} {\bibfnamefont {R.}~\bibnamefont {Orlando}},\ }\href {\doibase
  10.1103/PhysRevB.89.045103} {\bibfield  {journal} {\bibinfo  {journal} {Phys.
  Rev. B}\ }\textbf {\bibinfo {volume} {89}},\ \bibinfo {pages} {045103}
  (\bibinfo {year} {2014})}\BibitemShut {NoStop}%
\bibitem [{\citenamefont {Heifets}\ \emph {et~al.}(2015)\citenamefont
  {Heifets}, \citenamefont {Kotomin}, \citenamefont {Bagaturyants},\ and\
  \citenamefont {Maier}}]{Heifets2015}%
  \BibitemOpen
  \bibfield  {author} {\bibinfo {author} {\bibfnamefont {E.}~\bibnamefont
  {Heifets}}, \bibinfo {author} {\bibfnamefont {E.~A.}\ \bibnamefont
  {Kotomin}}, \bibinfo {author} {\bibfnamefont {A.~A.}\ \bibnamefont
  {Bagaturyants}}, \ and\ \bibinfo {author} {\bibfnamefont {J.}~\bibnamefont
  {Maier}},\ }\href {\doibase 10.1021/acs.jpclett.5b01071} {\bibfield
  {journal} {\bibinfo  {journal} {The Journal of Physical Chemistry Letters}\
  }\textbf {\bibinfo {volume} {6}},\ \bibinfo {pages} {2847} (\bibinfo {year}
  {2015})}\BibitemShut {NoStop}%
\bibitem [{\citenamefont {Erba}\ \emph {et~al.}(2013)\citenamefont {Erba},
  \citenamefont {El-Kelany}, \citenamefont {Ferrero}, \citenamefont
  {Baraille},\ and\ \citenamefont {R\'erat}}]{Erba2013}%
  \BibitemOpen
  \bibfield  {author} {\bibinfo {author} {\bibfnamefont {A.}~\bibnamefont
  {Erba}}, \bibinfo {author} {\bibfnamefont {K.~E.}\ \bibnamefont {El-Kelany}},
  \bibinfo {author} {\bibfnamefont {M.}~\bibnamefont {Ferrero}}, \bibinfo
  {author} {\bibfnamefont {I.}~\bibnamefont {Baraille}}, \ and\ \bibinfo
  {author} {\bibfnamefont {M.}~\bibnamefont {R\'erat}},\ }\href {\doibase
  10.1103/PhysRevB.88.035102} {\bibfield  {journal} {\bibinfo  {journal} {Phys.
  Rev. B}\ }\textbf {\bibinfo {volume} {88}},\ \bibinfo {pages} {035102}
  (\bibinfo {year} {2013})}\BibitemShut {NoStop}%
\bibitem [{\citenamefont {Cherair}\ \emph {et~al.}(2017)\citenamefont
  {Cherair}, \citenamefont {Iles}, \citenamefont {Rabahi},\ and\ \citenamefont
  {Kellou}}]{CHERAIR2016}%
  \BibitemOpen
  \bibfield  {author} {\bibinfo {author} {\bibfnamefont {I.}~\bibnamefont
  {Cherair}}, \bibinfo {author} {\bibfnamefont {N.}~\bibnamefont {Iles}},
  \bibinfo {author} {\bibfnamefont {L.}~\bibnamefont {Rabahi}}, \ and\ \bibinfo
  {author} {\bibfnamefont {A.}~\bibnamefont {Kellou}},\ }\href {\doibase
  10.1016/j.commatsci.2016.10.018} {\bibfield  {journal} {\bibinfo  {journal}
  {Computational Materials Science}\ }\textbf {\bibinfo {volume} {126}},\
  \bibinfo {pages} {491 } (\bibinfo {year} {2017})}\BibitemShut {NoStop}%
\bibitem [{\citenamefont {Noura}(2014)}]{Hamdad2014}%
  \BibitemOpen
  \bibfield  {author} {\bibinfo {author} {\bibfnamefont {H.}~\bibnamefont
  {Noura}},\ }\href {\doibase 10.1016/j.spmi.2014.10.004} {\bibfield  {journal}
  {\bibinfo  {journal} {Superlattices and Microstructures}\ }\textbf {\bibinfo
  {volume} {76}},\ \bibinfo {pages} {425 } (\bibinfo {year}
  {2014})}\BibitemShut {NoStop}%
\bibitem [{\citenamefont {Verma}\ and\ \citenamefont
  {Kumar}(2012)}]{Bulk-modulus}%
  \BibitemOpen
  \bibfield  {author} {\bibinfo {author} {\bibfnamefont {A.}~\bibnamefont
  {Verma}}\ and\ \bibinfo {author} {\bibfnamefont {A.}~\bibnamefont {Kumar}},\
  }\href {\doibase 10.1016/j.jallcom.2012.07.027} {\bibfield  {journal}
  {\bibinfo  {journal} {Journal of Alloys and Compounds}\ }\textbf {\bibinfo
  {volume} {541}},\ \bibinfo {pages} {210 } (\bibinfo {year}
  {2012})}\BibitemShut {NoStop}%
\bibitem [{\citenamefont {Li}\ \emph {et~al.}(2013)\citenamefont {Li},
  \citenamefont {Feng}, \citenamefont {Jing}, \citenamefont {Hong},
  \citenamefont {Cao},\ and\ \citenamefont {Zhang}}]{LiTongwei2013}%
  \BibitemOpen
  \bibfield  {author} {\bibinfo {author} {\bibfnamefont {T.}~\bibnamefont
  {Li}}, \bibinfo {author} {\bibfnamefont {Z.}~\bibnamefont {Feng}}, \bibinfo
  {author} {\bibfnamefont {C.}~\bibnamefont {Jing}}, \bibinfo {author}
  {\bibfnamefont {F.}~\bibnamefont {Hong}}, \bibinfo {author} {\bibfnamefont
  {S.}~\bibnamefont {Cao}}, \ and\ \bibinfo {author} {\bibfnamefont
  {J.}~\bibnamefont {Zhang}},\ }\href {\doibase 10.1140/epjb/e2013-40005-8}
  {\bibfield  {journal} {\bibinfo  {journal} {The European Physical Journal B}\
  }\textbf {\bibinfo {volume} {86}},\ \bibinfo {pages} {1} (\bibinfo {year}
  {2013})}\BibitemShut {NoStop}%
\bibitem [{\citenamefont {Boukhvalov}\ and\ \citenamefont
  {Solovyev}(2010)}]{Boukvalov2010}%
  \BibitemOpen
  \bibfield  {author} {\bibinfo {author} {\bibfnamefont {D.~W.}\ \bibnamefont
  {Boukhvalov}}\ and\ \bibinfo {author} {\bibfnamefont {I.~V.}\ \bibnamefont
  {Solovyev}},\ }\href {\doibase 10.1103/PhysRevB.82.245101} {\bibfield
  {journal} {\bibinfo  {journal} {Phys. Rev. B}\ }\textbf {\bibinfo {volume}
  {82}},\ \bibinfo {pages} {245101} (\bibinfo {year} {2010})}\BibitemShut
  {NoStop}%
\bibitem [{\citenamefont {He}\ and\ \citenamefont {Franchini}(2012)}]{He2012}%
  \BibitemOpen
  \bibfield  {author} {\bibinfo {author} {\bibfnamefont {J.}~\bibnamefont
  {He}}\ and\ \bibinfo {author} {\bibfnamefont {C.}~\bibnamefont {Franchini}},\
  }\href {\doibase 10.1103/PhysRevB.86.235117} {\bibfield  {journal} {\bibinfo
  {journal} {Phys. Rev. B}\ }\textbf {\bibinfo {volume} {86}},\ \bibinfo
  {pages} {235117} (\bibinfo {year} {2012})}\BibitemShut {NoStop}%
\bibitem [{\citenamefont {Sawada}\ \emph {et~al.}(1997)\citenamefont {Sawada},
  \citenamefont {Morikawa}, \citenamefont {Terakura},\ and\ \citenamefont
  {Hamada}}]{Sawada1997}%
  \BibitemOpen
  \bibfield  {author} {\bibinfo {author} {\bibfnamefont {H.}~\bibnamefont
  {Sawada}}, \bibinfo {author} {\bibfnamefont {Y.}~\bibnamefont {Morikawa}},
  \bibinfo {author} {\bibfnamefont {K.}~\bibnamefont {Terakura}}, \ and\
  \bibinfo {author} {\bibfnamefont {N.}~\bibnamefont {Hamada}},\ }\href
  {\doibase 10.1103/PhysRevB.56.12154} {\bibfield  {journal} {\bibinfo
  {journal} {Physical Review B}\ }\textbf {\bibinfo {volume} {56}},\ \bibinfo
  {pages} {12154} (\bibinfo {year} {1997})}\BibitemShut {NoStop}%
\bibitem [{\citenamefont {Mellan}\ \emph {et~al.}(2015)\citenamefont {Mellan},
  \citenamefont {Cor\`a}, \citenamefont {Grau-Crespo},\ and\ \citenamefont
  {Ismail-Beigi}}]{mellan2015}%
  \BibitemOpen
  \bibfield  {author} {\bibinfo {author} {\bibfnamefont {T.~A.}\ \bibnamefont
  {Mellan}}, \bibinfo {author} {\bibfnamefont {F.}~\bibnamefont {Cor\`a}},
  \bibinfo {author} {\bibfnamefont {R.}~\bibnamefont {Grau-Crespo}}, \ and\
  \bibinfo {author} {\bibfnamefont {S.}~\bibnamefont {Ismail-Beigi}},\ }\href
  {\doibase 10.1103/PhysRevB.92.085151} {\bibfield  {journal} {\bibinfo
  {journal} {Phys. Rev. B}\ }\textbf {\bibinfo {volume} {92}},\ \bibinfo
  {pages} {085151} (\bibinfo {year} {2015})}\BibitemShut {NoStop}%
\bibitem [{\citenamefont {Woodward}(1997{\natexlab{a}})}]{woodward1997}%
  \BibitemOpen
  \bibfield  {author} {\bibinfo {author} {\bibfnamefont {P.~M.}\ \bibnamefont
  {Woodward}},\ }\href@noop {} {\bibfield  {journal} {\bibinfo  {journal} {Acta
  Cryst. B}\ }\textbf {\bibinfo {volume} {53}},\ \bibinfo {pages} {32}
  (\bibinfo {year} {1997}{\natexlab{a}})}\BibitemShut {NoStop}%
\bibitem [{\citenamefont {Woodward}(1997{\natexlab{b}})}]{woodward1997b}%
  \BibitemOpen
  \bibfield  {author} {\bibinfo {author} {\bibfnamefont {P.~M.}\ \bibnamefont
  {Woodward}},\ }\href@noop {} {\bibfield  {journal} {\bibinfo  {journal} {Acta
  Cryst. B}\ }\textbf {\bibinfo {volume} {53}},\ \bibinfo {pages} {44}
  (\bibinfo {year} {1997}{\natexlab{b}})}\BibitemShut {NoStop}%
\bibitem [{\citenamefont {Goldschmidt}(1926)}]{goldschmidt1926}%
  \BibitemOpen
  \bibfield  {author} {\bibinfo {author} {\bibfnamefont {V.~M.}\ \bibnamefont
  {Goldschmidt}},\ }\href {\doibase 10.1007/bf01507527} {\bibfield  {journal}
  {\bibinfo  {journal} {Die Naturwissenschaften}\ }\textbf {\bibinfo {volume}
  {21}},\ \bibinfo {pages} {477} (\bibinfo {year} {1926})}\BibitemShut
  {NoStop}%
\bibitem [{\citenamefont {Balachandran}\ and\ \citenamefont
  {Rondinelli}(2013)}]{Breathing}%
  \BibitemOpen
  \bibfield  {author} {\bibinfo {author} {\bibfnamefont {P.~V.}\ \bibnamefont
  {Balachandran}}\ and\ \bibinfo {author} {\bibfnamefont {J.~M.}\ \bibnamefont
  {Rondinelli}},\ }\href {\doibase 10.1103/PhysRevB.88.054101} {\bibfield
  {journal} {\bibinfo  {journal} {Phys. Rev. B}\ }\textbf {\bibinfo {volume}
  {88}},\ \bibinfo {pages} {054101} (\bibinfo {year} {2013})}\BibitemShut
  {NoStop}%
\bibitem [{\citenamefont {Maznichenko}\ \emph {et~al.}(2016)\citenamefont
  {Maznichenko}, \citenamefont {Ostanin}, \citenamefont {Bekenov},
  \citenamefont {Antonov}, \citenamefont {Mertig},\ and\ \citenamefont
  {Ernst}}]{Maznichenko2016}%
  \BibitemOpen
  \bibfield  {author} {\bibinfo {author} {\bibfnamefont {I.~V.}\ \bibnamefont
  {Maznichenko}}, \bibinfo {author} {\bibfnamefont {S.}~\bibnamefont
  {Ostanin}}, \bibinfo {author} {\bibfnamefont {L.~V.}\ \bibnamefont
  {Bekenov}}, \bibinfo {author} {\bibfnamefont {V.~N.}\ \bibnamefont
  {Antonov}}, \bibinfo {author} {\bibfnamefont {I.}~\bibnamefont {Mertig}}, \
  and\ \bibinfo {author} {\bibfnamefont {A.}~\bibnamefont {Ernst}},\ }\href
  {\doibase 10.1103/PhysRevB.93.024411} {\bibfield  {journal} {\bibinfo
  {journal} {Phys. Rev. B}\ }\textbf {\bibinfo {volume} {93}},\ \bibinfo
  {pages} {024411} (\bibinfo {year} {2016})}\BibitemShut {NoStop}%
\bibitem [{\citenamefont {Shi}\ and\ \citenamefont {et~al.}(2013)}]{shi2014}%
  \BibitemOpen
  \bibfield  {author} {\bibinfo {author} {\bibfnamefont {Y.}~\bibnamefont
  {Shi}}\ and\ \bibinfo {author} {\bibnamefont {et~al.}},\ }\href {\doibase
  10.1038/nmat3754} {\bibfield  {journal} {\bibinfo  {journal} {Nature
  Materials}\ }\textbf {\bibinfo {volume} {12}},\ \bibinfo {pages} {1024}
  (\bibinfo {year} {2013})}\BibitemShut {NoStop}%
\bibitem [{\citenamefont {Garcia-Castro}\ \emph {et~al.}(2014)\citenamefont
  {Garcia-Castro}, \citenamefont {Spaldin}, \citenamefont {Romero},\ and\
  \citenamefont {Bousquet}}]{garcia-castro2014}%
  \BibitemOpen
  \bibfield  {author} {\bibinfo {author} {\bibfnamefont {A.~C.}\ \bibnamefont
  {Garcia-Castro}}, \bibinfo {author} {\bibfnamefont {N.~A.}\ \bibnamefont
  {Spaldin}}, \bibinfo {author} {\bibfnamefont {A.~H.}\ \bibnamefont {Romero}},
  \ and\ \bibinfo {author} {\bibfnamefont {E.}~\bibnamefont {Bousquet}},\
  }\href {\doibase 10.1103/PhysRevB.89.104107} {\bibfield  {journal} {\bibinfo
  {journal} {Phys. Rev. B}\ }\textbf {\bibinfo {volume} {89}},\ \bibinfo
  {pages} {104107} (\bibinfo {year} {2014})}\BibitemShut {NoStop}%
\bibitem [{\citenamefont {Rahman}\ and\ \citenamefont
  {Sarwar}(2016)}]{Gul2016-strain}%
  \BibitemOpen
  \bibfield  {author} {\bibinfo {author} {\bibfnamefont {G.}~\bibnamefont
  {Rahman}}\ and\ \bibinfo {author} {\bibfnamefont {S.}~\bibnamefont
  {Sarwar}},\ }\href {\doibase 10.1002/pssb.201552417} {\bibfield  {journal}
  {\bibinfo  {journal} {physica status solidi (b)}\ }\textbf {\bibinfo {volume}
  {253}},\ \bibinfo {pages} {329} (\bibinfo {year} {2016})}\BibitemShut
  {NoStop}%
\bibitem [{\citenamefont {Tsuyama}\ \emph {et~al.}(2016)\citenamefont
  {Tsuyama}, \citenamefont {Chakraverty}, \citenamefont {Macke}, \citenamefont
  {Pontius}, \citenamefont {Sch\"u\ss{}ler-Langeheine}, \citenamefont {Hwang},
  \citenamefont {Tokura},\ and\ \citenamefont {Wadati}}]{Tsuyama2016}%
  \BibitemOpen
  \bibfield  {author} {\bibinfo {author} {\bibfnamefont {T.}~\bibnamefont
  {Tsuyama}}, \bibinfo {author} {\bibfnamefont {S.}~\bibnamefont
  {Chakraverty}}, \bibinfo {author} {\bibfnamefont {S.}~\bibnamefont {Macke}},
  \bibinfo {author} {\bibfnamefont {N.}~\bibnamefont {Pontius}}, \bibinfo
  {author} {\bibfnamefont {C.}~\bibnamefont {Sch\"u\ss{}ler-Langeheine}},
  \bibinfo {author} {\bibfnamefont {H.~Y.}\ \bibnamefont {Hwang}}, \bibinfo
  {author} {\bibfnamefont {Y.}~\bibnamefont {Tokura}}, \ and\ \bibinfo {author}
  {\bibfnamefont {H.}~\bibnamefont {Wadati}},\ }\href {\doibase
  10.1103/PhysRevLett.116.256402} {\bibfield  {journal} {\bibinfo  {journal}
  {Phys. Rev. Lett.}\ }\textbf {\bibinfo {volume} {116}},\ \bibinfo {pages}
  {256402} (\bibinfo {year} {2016})}\BibitemShut {NoStop}%
\bibitem [{\citenamefont {Jeng}\ and\ \citenamefont {Guo}(2002)}]{Jeng2002}%
  \BibitemOpen
  \bibfield  {author} {\bibinfo {author} {\bibfnamefont {H.-T.}\ \bibnamefont
  {Jeng}}\ and\ \bibinfo {author} {\bibfnamefont {G.}~\bibnamefont {Guo}},\
  }\href {\doibase http://dx.doi.org/10.1103/PhysRevB.65.094429} {\bibfield
  {journal} {\bibinfo  {journal} {Physical review B}\ }\textbf {\bibinfo
  {volume} {65}},\ \bibinfo {pages} {094429} (\bibinfo {year}
  {2002})}\BibitemShut {NoStop}%
\bibitem [{\citenamefont {Fri\'{a}k}\ \emph {et~al.}(2007)\citenamefont
  {Fri\'{a}k}, \citenamefont {Schindlmayr},\ and\ \citenamefont
  {Scheffler}}]{Friak2007}%
  \BibitemOpen
  \bibfield  {author} {\bibinfo {author} {\bibfnamefont {M.}~\bibnamefont
  {Fri\'{a}k}}, \bibinfo {author} {\bibfnamefont {A.}~\bibnamefont
  {Schindlmayr}}, \ and\ \bibinfo {author} {\bibfnamefont {M.}~\bibnamefont
  {Scheffler}},\ }\href {\doibase 10.1088/1367-2630/9/1/005} {\bibfield
  {journal} {\bibinfo  {journal} {New Journal of Physics}\ }\textbf {\bibinfo
  {volume} {9}},\ \bibinfo {pages} {5} (\bibinfo {year} {2007})}\BibitemShut
  {NoStop}%
\bibitem [{\citenamefont {Mizokawa}\ \emph {et~al.}(1999)\citenamefont
  {Mizokawa}, \citenamefont {Khomskii},\ and\ \citenamefont
  {Sawatzky}}]{Mizokawa1999}%
  \BibitemOpen
  \bibfield  {author} {\bibinfo {author} {\bibfnamefont {T.}~\bibnamefont
  {Mizokawa}}, \bibinfo {author} {\bibfnamefont {D.~I.}\ \bibnamefont
  {Khomskii}}, \ and\ \bibinfo {author} {\bibfnamefont {G.~A.}\ \bibnamefont
  {Sawatzky}},\ }\href {\doibase 10.1103/PhysRevB.60.7309} {\bibfield
  {journal} {\bibinfo  {journal} {Phys. Rev. B}\ }\textbf {\bibinfo {volume}
  {60}},\ \bibinfo {pages} {7309} (\bibinfo {year} {1999})}\BibitemShut
  {NoStop}%
\bibitem [{\citenamefont {Carpenter}\ and\ \citenamefont
  {Howard}(2009)}]{Carpenter2009}%
  \BibitemOpen
  \bibfield  {author} {\bibinfo {author} {\bibfnamefont {M.~A.}\ \bibnamefont
  {Carpenter}}\ and\ \bibinfo {author} {\bibfnamefont {C.~J.}\ \bibnamefont
  {Howard}},\ }\href {\doibase 10.1107/S0108768109000974} {\bibfield  {journal}
  {\bibinfo  {journal} {Acta Crystallographica Section B}\ }\textbf {\bibinfo
  {volume} {65}},\ \bibinfo {pages} {134} (\bibinfo {year} {2009})}\BibitemShut
  {NoStop}%
\bibitem [{\citenamefont {Song}\ and\ \citenamefont {Zhang}(2016)}]{Song2016}%
  \BibitemOpen
  \bibfield  {author} {\bibinfo {author} {\bibfnamefont {G.}~\bibnamefont
  {Song}}\ and\ \bibinfo {author} {\bibfnamefont {W.}~\bibnamefont {Zhang}},\
  }\href {\doibase 10.1103/PhysRevB.94.064409} {\bibfield  {journal} {\bibinfo
  {journal} {Phys. Rev. B}\ }\textbf {\bibinfo {volume} {94}},\ \bibinfo
  {pages} {064409} (\bibinfo {year} {2016})}\BibitemShut {NoStop}%
\end{thebibliography}
\end{document}